\documentclass[fleqn,usenatbib,useAMS]{mnras}

\usepackage{graphicx}	
\usepackage{amsmath}	
\usepackage{amssymb}	
\usepackage{multicol}        
\usepackage{bm}		
\usepackage{pdflscape}	
\usepackage{csvsimple}

\usepackage[T1]{fontenc}
\usepackage{ae,aecompl}

\usepackage{newtxtext,newtxmath}

\title[Timing noise in pulsars] {Diagnostics of timing noise in middle aged pulsars}

\author[Namkham et al.]
{Nakornping Namkham$^{1}$\thanks{E-mail: nakornping.namkham@gmail.com},
Phrudth Jaroenjittichai$^{2}$,
Simon Johnston$^{3}$
\\
$^{1}$Master's degree in Astronomy, Department of Physics and Materials, Faculty of Science, Chiang Mai University, Chiang Mai, Thailand, 50200\\
$^{2}$National Astronomical Research Institute of Thailand (Public Organization), Chiang Mai, Thailand, 50180\\
$^{3}$CSIRO Astronomy and Space Science, Australia Telescope National Facility, PO~Box~76, Epping NSW~1710, Australia\\
}
\date{Last updated; in original form}

\pubyear{2019}

\begin{document}
\label{firstpage}
\pagerange{\pageref{firstpage}--\pageref{lastpage}}
\maketitle

\begin{abstract}
Radio pulsars are often used as clocks in a wide variety of experiments. Imperfections in the clock, known as timing noise, have the potential to reduce the significance of, or even thwart e.g. the attempt to find a stochastic gravitational wave (GW) background. We measure the timing noise in a group of 129 mostly middle-aged pulsars (i.e. characterstic ages near 1~Myr) observed with the Parkes radio telescope on a monthly basis since 2014. We examine four different metrics for timing noise, but it remains unclear which, if any, provides the best determination. In spite of this, it is evident that these pulsars have significantly less timing noise than their younger counterparts, but significantly more than the (much older) millisecond pulsars (MSPs). As with previous authors, we find a strong correlation between timing noise and the pulsar spin-down rate, $\dot{\nu}$. However, for a given $\dot{\nu}$ there is a spread of about a factor 30 in the strength of the timing noise likely indicating that nuclear conditions in the interior of the stars differs between objects. We briefly comment on the implications for GW detection through pulsar timing arrays as the level of timing noise in MSPs may be less than predicted.
\end{abstract}

\begin{keywords}
pulsars
\end{keywords}


\section{Introduction}
The rotation of a pulsar, coupled with a narrow beam of radio emission emanating from a small region around the magnetic poles, means that an Earth-based observer sees a regular pulse of radio emission once per rotation. This clock-like behaviour of pulsars has a wide variety of applications, from the testing of theories of gravity
(e.g. \citealt{ksm+06,sfc+18,cck+18})
to possible detection of the stochastic background of gravitational waves \citep{srl+15,abb+18}. However, the clocks are far from perfect even after taking into account braking torques and other effects and it is known that the young pulsars have imperfections in their rotation rate which are much larger than those of the older, millisecond pulsars \citep{sc10}. 
These are attributed to irregularities in the rotation of the neutron star itself, likely caused by the imperfect coupling between the crust of the star and its superfluid interior \citep{jones90,ml14} or due to changes in the conditions in the magnetosphere, either intrinsically (e.g. \citealt{lhk+10,bkj16}) or extrinsically (\citealt{bkb+14}) driven.

A number of different diagnostic tools has been used over the years to measure the timing noise in pulsars; we consider four such diagnostics here. Following pioneering work by \citet{ch80} who first derived a quantitative measure of timing noise, \citet{antt94} measured the timing noise in a large sample of pulsars. They defined the metric $\Delta_8$
\begin{equation}
\label{delta8}
    \Delta_8 = \frac{|\ddot{\nu}|}{6\nu} T^3
\end{equation}
where $\nu$ is the pulsar's spin frequency and $T$ is the total observing span which in the case of the observational sample of \citet{antt94} was $10^8$~s. However, this metric cannot be used to compare pulsars with different observation lengths, $T$, as in general $\ddot{\nu}$ is a complex function of $T$. 
\citet{hlk10} used a metric, $\sigma_z$, originally from \citet{mte97}, based on the Allan variance of terrestrial clocks:
\begin{equation}
\label{sigmaz}
    \sigma_z(T) = \frac{T^2}{2 \sqrt{5}}\langle\ddot{\nu}^2\rangle^{1/2}
\end{equation}
which uses the average magnitude of $\ddot{\nu}$ over short time spans, $T$. Again, $\sigma_z$ depends on the time-span. As \citet{hlk10} showed, if the power spectral density of the timing residuals is itself a power-law with spectral index $\alpha_{\rm R}$ then
\begin{equation}
\label{zpt}
    \sigma_z(T) \propto T^{(\alpha_{\rm R}+3)/2}
\end{equation}
\begin{table*}
\centering
\caption{Basic parameters for the 129 pulsars used in this work}
    \label{params}
\begin{multicols}{2}
    \begin{tabular}{lccccc}
	Name & \multicolumn{1}{|p{0.6cm}|}{\centering $\nu$ \\ (Hz)} & \multicolumn{1}{|p{0.6cm}|}{\centering $\log|\dot{\nu}|$ \\ (/s$^{-2}$)} & \multicolumn{1}{|p{0.6cm}|}{\centering $\log\dot{E}$ \\ (/ergs$^{-1}$)} & \multicolumn{1}{|p{0.6cm}|}{\centering $\log\tau_c$ \\ (/yr)} & \multicolumn{1}{|p{0.6cm}|}{\centering $\log B$ \\ (/G)} \\\hline
	J0034  $  -  $  0721  &  1.06  & $  -15.34  $ &  31.28  &  7.56  &  11.79
 \\
J0051  $  +  $  0423  &  2.82  & $  -16.30  $ &  30.73  &  8.97  &  10.67
 \\
J0108  $  -  $  1431  &  1.24  & $  -15.92  $ &  30.76  &  8.22  &  11.40
 \\
J0134  $  -  $  2937  &  7.30  & $  -14.38  $ &  33.08  &  7.44  &  11.02
 \\
J0152  $  -  $  1637  &  1.20  & $  -14.73  $ &  31.95  &  7.01  &  12.02
 \\
J0206  $  -  $  4028  &  1.59  & $  -14.52  $ &  32.28  &  6.92  &  11.94
 \\
J0255  $  -  $  5304  &  2.23  & $  -15.82  $ &  31.13  &  8.37  &  11.07
 \\
J0343  $  -  $  3000  &  0.39  & $  -17.00  $ &  29.28  &  8.69  &  11.67
 \\
J0401  $  -  $  7608  &  1.83  & $  -14.29  $ &  32.57  &  6.75  &  11.96
 \\
J0448  $  -  $  2749  &  2.22  & $  -15.13  $ &  31.81  &  7.68  &  11.41
 \\
J0452  $  -  $  1759  &  1.82  & $  -13.72  $ &  33.14  &  6.18  &  12.25
 \\
J0525  $  +  $  1115  &  2.82  & $  -15.23  $ &  31.82  &  7.88  &  11.21
 \\
J0536  $  -  $  7543  &  0.80  & $  -15.43  $ &  31.07  &  7.53  &  11.93
 \\
J0601  $  -  $  0527  &  2.53  & $  -14.08  $ &  32.92  &  6.68  &  11.86
 \\
J0624  $  -  $  0424  &  0.96  & $  -15.11  $ &  31.46  &  7.30  &  11.97
 \\
J0630  $  -  $  2834  &  0.80  & $  -14.33  $ &  32.17  &  6.44  &  12.48
 \\
J0729  $  -  $  1836  &  1.96  & $  -13.13  $ &  33.75  &  5.63  &  12.49
 \\
J0738  $  -  $  4042  &  2.67  & $  -14.01  $ &  33.01  &  6.63  &  11.86
 \\
J0809  $  -  $  4753  &  1.83  & $  -13.99  $ &  32.87  &  6.45  &  12.11
 \\
J0820  $  -  $  1350  &  0.81  & $  -14.86  $ &  31.64  &  6.97  &  12.21
 \\
J0820  $  -  $  4114  &  1.83  & $  -16.22  $ &  30.61  &  8.71  &  10.98
 \\
J0837  $  -  $  4135  &  1.33  & $  -14.20  $ &  32.52  &  6.52  &  12.21
 \\
J0904  $  -  $  7459  &  1.82  & $  -14.82  $ &  32.04  &  7.27  &  11.70
 \\
J0907  $  -  $  5157  &  3.94  & $  -13.54  $ &  33.65  &  6.34  &  11.84
 \\
J0924  $  -  $  5814  &  1.35  & $  -14.07  $ &  32.66  &  6.40  &  12.27
 \\
J0942  $  -  $  5552  &  1.51  & $  -13.20  $ &  33.57  &  5.58  &  12.63
 \\
J1001  $  -  $  5507  &  0.70  & $  -13.59  $ &  32.85  &  5.63  &  12.94
 \\
J1034  $  -  $  3224  &  0.87  & $  -15.77  $ &  30.78  &  7.90  &  11.71
 \\
J1038  $  -  $  5831  &  1.51  & $  -14.56  $ &  32.21  &  6.94  &  11.95
 \\
J1047  $  -  $  6709  &  5.04  & $  -13.37  $ &  33.93  &  6.27  &  11.76
 \\
J1055  $  -  $  6022  &  1.06  & $  -12.99  $ &  33.63  &  5.21  &  12.97
 \\
J1056  $  -  $  6258  &  2.37  & $  -13.69  $ &  33.28  &  6.27  &  12.09
 \\
J1136  $  -  $  5525  &  2.74  & $  -13.23  $ &  33.81  &  5.86  &  12.23
 \\
J1146  $  -  $  6030  &  3.66  & $  -13.62  $ &  33.54  &  6.38  &  11.84
 \\
J1157  $  -  $  6224  &  2.50  & $  -13.61  $ &  33.38  &  6.21  &  12.10
 \\
J1243  $  -  $  6423  &  2.57  & $  -13.52  $ &  33.48  &  6.13  &  12.12
 \\
J1306  $  -  $  6617  &  2.11  & $  -13.55  $ &  33.37  &  6.08  &  12.23
 \\
J1317  $  -  $  6302  &  3.83  & $  -14.83  $ &  32.35  &  7.61  &  11.21
 \\
J1319  $  -  $  6056  &  3.52  & $  -13.72  $ &  33.42  &  6.47  &  11.82
 \\
J1326  $  -  $  5859  &  2.09  & $  -13.82  $ &  33.10  &  6.34  &  12.11
 \\
J1326  $  -  $  6408  &  1.26  & $  -14.31  $ &  32.39  &  6.61  &  12.19
 \\
J1326  $  -  $  6700  &  1.84  & $  -13.74  $ &  33.12  &  6.20  &  12.23
 \\
J1327  $  -  $  6222  &  1.89  & $  -13.15  $ &  33.72  &  5.63  &  12.51
 \\
J1327  $  -  $  6301  &  5.09  & $  -13.40  $ &  33.90  &  6.31  &  11.74
 \\
J1328  $  -  $  4357  &  1.88  & $  -13.99  $ &  32.88  &  6.46  &  12.10
 \\
J1338  $  -  $  6204  &  0.81  & $  -14.05  $ &  32.46  &  6.15  &  12.62
 \\
J1340  $  -  $  6456  &  2.64  & $  -13.44  $ &  33.57  &  6.07  &  12.14
 \\
J1352  $  -  $  6803  &  1.59  & $  -14.50  $ &  32.30  &  6.90  &  11.95
 \\
J1356  $  -  $  5521  &  1.97  & $  -14.54  $ &  32.35  &  7.03  &  11.79
 \\
J1357  $  -  $  62  &  2.19  & $  -14.54  $ &  33.79  &  5.69  &  12.41
 \\
J1401  $  -  $  6357  &  1.19  & $  -13.15  $ &  33.00  &  5.94  &  12.56
 \\
J1418  $  -  $  3921  &  0.91  & $  -13.67  $ &  31.43  &  7.29  &  12.00
 \\
J1424  $  -  $  5822  &  2.73  & $  -15.13  $ &  33.50  &  6.17  &  12.08
 \\
J1428  $  -  $  5530  &  1.75  & $  -13.53  $ &  32.65  &  6.63  &  12.04
 \\
J1430  $  -  $  6623  &  1.27  & $  -14.19  $ &  32.36  &  6.65  &  12.17
 \\
J1435  $  -  $  5954  &  2.11  & $  -14.35  $ &  32.76  &  6.68  &  11.93
 \\
J1456  $  -  $  6843  &  3.80  & $  -14.16  $ &  32.33  &  7.62  &  11.21
 \\
J1522  $  -  $  5829  &  2.53  & $  -14.84  $ &  33.11  &  6.49  &  11.95
 \\
J1534  $  -  $  5334  &  0.73  & $  -13.89  $ &  31.34  &  7.18  &  12.15
 \\
J1534  $  -  $  5405  &  3.45  & $  -15.12  $ &  33.40  &  6.47  &  11.82
 \\
J1536  $  -  $  5433  &  1.13  & $  -13.74  $ &  32.00  &  6.90  &  12.09 \\

    \end{tabular}
    
    \begin{tabular}{lccccc}
	Name & \multicolumn{1}{|p{0.6cm}|}{\centering $\nu$ \\ (Hz)} & \multicolumn{1}{|p{0.6cm}|}{\centering $\log|\dot{\nu}|$ \\ (/s$^{-2}$)} & \multicolumn{1}{|p{0.6cm}|}{\centering $\log\dot{E}$ \\ (/ergs$^{-1}$)} & \multicolumn{1}{|p{0.6cm}|}{\centering $\log\tau_c$ \\ (/yr)} & \multicolumn{1}{|p{0.6cm}|}{\centering $\log B$ \\ (/G)} \\\hline
	J1544  $  -  $  5308  &  5.60  & $  -14.65  $ &  32.63  &  7.67  &  11.02
 \\
J1555  $  -  $  3134  &  1.93  & $  -14.72  $ &  31.25  &  8.12  &  11.25
 \\
J1557  $  -  $  4258  &  3.04  & $  -15.64  $ &  32.56  &  7.21  &  11.51
 \\
J1559  $  -  $  4438  &  3.89  & $  -14.52  $ &  33.38  &  6.60  &  11.71
 \\
J1604  $  -  $  4909  &  3.05  & $  -13.81  $ &  33.00  &  6.76  &  11.73
 \\
J1605  $  -  $  5257  &  1.52  & $  -14.08  $ &  31.55  &  7.61  &  11.61
 \\
J1613  $  -  $  4714  &  2.62  & $  -15.23  $ &  32.65  &  6.98  &  11.69
 \\
J1623  $  -  $  4256  &  2.74  & $  -14.36  $ &  32.88  &  6.79  &  11.76
 \\
J1626  $  -  $  4537  &  2.70  & $  -14.16  $ &  33.81  &  5.85  &  12.24
 \\
J1630  $  -  $  4733  &  1.74  & $  -13.22  $ &  33.66  &  5.61  &  12.55
 \\
J1632  $  -  $  4621  &  0.59  & $  -13.17  $ &  32.78  &  5.55  &  13.06
 \\
J1633  $  -  $  4453  &  2.29  & $  -13.58  $ &  33.47  &  6.05  &  12.22
 \\
J1633  $  -  $  5015  &  2.84  & $  -13.49  $ &  33.54  &  6.17  &  12.06
 \\
J1646  $  -  $  6831  &  0.56  & $  -13.01  $ &  31.08  &  7.22  &  12.24
 \\
J1649  $  -  $  3805  &  3.82  & $  -14.17  $ &  31.89  &  8.07  &  10.98
 \\
J1651  $  -  $  4246  &  1.18  & $  -15.29  $ &  32.45  &  6.49  &  12.28
 \\
J1651  $  -  $  5222  &  1.57  & $  -14.22  $ &  32.45  &  6.74  &  12.03
 \\
J1652  $  -  $  2404  &  0.59  & $  -14.34  $ &  31.40  &  6.93  &  12.36
 \\
J1653  $  -  $  3838  &  3.28  & $  -14.97  $ &  33.58  &  6.24  &  11.96
 \\
J1653  $  -  $  4249  &  1.63  & $  -13.53  $ &  32.92  &  6.30  &  12.24
 \\
J1700  $  -  $  3312  &  0.74  & $  -13.89  $ &  31.86  &  6.66  &  12.40
 \\
J1701  $  -  $  3726  &  0.41  & $  -14.60  $ &  31.47  &  6.54  &  12.72
 \\
J1703  $  -  $  3241  &  0.83  & $  -14.38  $ &  31.17  &  7.46  &  11.95
 \\
J1703  $  -  $  4851  &  0.72  & $  -15.34  $ &  31.86  &  6.64  &  12.42
 \\
J1705  $  -  $  1906  &  3.34  & $  -14.59  $ &  33.79  &  6.06  &  12.05
 \\
J1705  $  -  $  3423  &  3.92  & $  -13.33  $ &  33.40  &  6.58  &  11.72
 \\
J1707  $  -  $  4053  &  1.72  & $  -13.79  $ &  32.59  &  6.68  &  12.02
 \\
J1707  $  -  $  4729  &  3.75  & $  -14.24  $ &  33.51  &  6.43  &  11.81
 \\
J1708  $  -  $  3426  &  1.44  & $  -13.66  $ &  32.70  &  6.42  &  12.23
 \\
J1709  $  -  $  1640  &  1.53  & $  -14.06  $ &  32.96  &  6.21  &  12.31
 \\
J1715  $  -  $  4034  &  0.48  & $  -13.82  $ &  31.14  &  7.02  &  12.40
 \\
J1717  $  -  $  3425  &  1.52  & $  -13.54  $ &  33.13  &  6.03  &  12.40
 \\
J1717  $  -  $  4054  &  1.13  & $  -13.65  $ &  32.32  &  6.58  &  12.26
 \\
J1717  $  -  $  5800  &  3.11  & $  -14.33  $ &  32.36  &  7.42  &  11.40
 \\
J1718  $  -  $  3718  &  0.30  & $  -12.85  $ &  33.22  &  4.52  &  13.87
 \\
J1719  $  -  $  4006  &  5.29  & $  -13.33  $ &  33.99  &  6.25  &  11.75
 \\
J1720  $  -  $  2933  &  1.61  & $  -14.72  $ &  32.09  &  7.12  &  11.83
 \\
J1722  $  -  $  3207  &  2.10  & $  -14.56  $ &  32.35  &  7.08  &  11.74
 \\
J1722  $  -  $  3632  &  2.51  & $  -13.55  $ &  33.44  &  6.15  &  12.13
 \\
J1727  $  -  $  2739  &  0.77  & $  -15.19  $ &  31.29  &  7.28  &  12.07
 \\
J1733  $  -  $  2228  &  1.15  & $  -16.15  $ &  30.50  &  8.41  &  11.33
 \\
J1738  $  -  $  3211  &  1.30  & $  -14.87  $ &  31.84  &  7.19  &  11.89
 \\
J1739  $  -  $  3131  &  1.89  & $  -13.23  $ &  33.64  &  5.70  &  12.47
 \\
J1741  $  -  $  2733  &  1.12  & $  -15.70  $ &  30.94  &  7.96  &  11.57
 \\
J1741  $  -  $  3016  &  0.53  & $  -14.60  $ &  31.72  &  6.52  &  12.62
 \\
J1741  $  -  $  3927  &  1.95  & $  -14.08  $ &  32.80  &  6.57  &  12.02
 \\
J1743  $  -  $  3150  &  0.41  & $  -13.68  $ &  32.53  &  5.50  &  13.23
 \\
J1749  $  -  $  3002  &  1.64  & $  -13.67  $ &  33.14  &  6.08  &  12.34
 \\
J1751  $  -  $  4657  &  1.35  & $  -14.63  $ &  32.10  &  6.96  &  11.99
 \\
J1752  $  -  $  2806  &  1.78  & $  -13.59  $ &  33.26  &  6.04  &  12.33
 \\
J1807  $  -  $  0847  &  6.11  & $  -14.97  $ &  32.42  &  7.95  &  10.84
 \\
J1816  $  -  $  2650  &  1.69  & $  -15.60  $ &  31.22  &  8.03  &  11.36
 \\
J1817  $  -  $  3618  &  2.58  & $  -13.83  $ &  33.17  &  6.44  &  11.96
 \\
J1820  $  -  $  0427  &  1.67  & $  -13.76  $ &  33.06  &  6.18  &  12.28
 \\
J1822  $  -  $  2256  &  0.53  & $  -15.41  $ &  30.91  &  7.34  &  12.20
 \\
J1822  $  -  $  4209  &  2.19  & $  -14.65  $ &  32.29  &  7.19  &  11.67
 \\
J1823  $  -  $  3106  &  3.52  & $  -13.44  $ &  33.70  &  6.19  &  11.96
 \\
J1829  $  -  $  1751  &  3.26  & $  -13.24  $ &  33.87  &  5.95  &  12.11
 \\
J1845  $  -  $  0434  &  2.05  & $  -13.32  $ &  33.59  &  5.83  &  12.37
 \\
J1848  $  -  $  0123  &  1.52  & $  -13.92  $ &  32.86  &  6.30  &  12.27 
 \\
J1852  $  -  $  0635  &  1.91  & $  -13.27  $ &  33.60  &  5.75  &  12.44
 \\

    \end{tabular}
\end{multicols}
\end{table*}
\begin{table}
\addtocounter{table}{-1}
    \centering
    \caption{Basic parameters for the 129 pulsars used in this work (continued)}
    \begin{tabular}{lccccc}
	Name & \multicolumn{1}{|p{0.6cm}|}{\centering $\nu$ \\ (Hz)} & \multicolumn{1}{|p{0.6cm}|}{\centering $\log|\dot{\nu}|$ \\ (/s$^{-2}$)} & \multicolumn{1}{|p{0.6cm}|}{\centering $\log\dot{E}$ \\ (/ergs$^{-1}$)} & \multicolumn{1}{|p{0.6cm}|}{\centering $\log\tau_c$ \\ (/yr)} & \multicolumn{1}{|p{0.6cm}|}{\centering $\log B$ \\ (/G)} \\\hline
	J1852  $  -  $  2610  &  2.97  & $  -15.10  $ &  31.97  &  7.77  &  11.24
 \\
J1900  $  -  $  2600  &  1.63  & $  -15.27  $ &  31.55  &  7.68  &  11.55
 \\
J1913  $  -  $  0440  &  1.21  & $  -14.24  $ &  32.44  &  6.52  &  12.25
 \\
J1941  $  -  $  2602  &  2.48  & $  -14.23  $ &  32.76  &  6.82  &  11.79
 \\
J2048  $  -  $  1616  &  0.51  & $  -14.55  $ &  31.76  &  6.45  &  12.67
 \\
J2330  $  -  $  2005  &  0.61  & $  -14.77  $ &  31.61  &  6.75  &  12.44
 \\
J2346  $  -  $  0609  &  0.85  & $  -15.00  $ &  31.53  &  7.12  &  12.11 \\

    \end{tabular}
\end{table}
We also use the braking index, $n$, as a diagnostic. $n$ is defined as
\begin{equation}
\label{bi}
    n = \frac{\nu \ddot{\nu}}{\dot{\nu}^2}
\end{equation}
For simple magnetic braking, one expects a constant $n=3$ over time, but this is only true for a small number of pulsars \citep{els17}. Instead, for many pulsars, the $\ddot{\nu}$ term is dominated by the timing noise resulting in large, time-dependent positive and negative values of $n$ \citep{jg99}. Note that equation~\ref{bi} does not contain $T$ explicitly, but as for $\Delta_8$, $\ddot{\nu}$ varies with time. However, if all pulsars are observed over the same time-span $T$, then effectively $\Delta_8 \propto \ddot{\nu}/\nu$ and $\sigma_z \propto \ddot{\nu}$ which can be compared directly with $n$.
\citet{sc10} claimed that using $\ddot{\nu}$ directly tended to underestimate the timing noise and preferred a simpler metric which depended on the rms of the residuals of the times-of-arrival (ToA) after fitting for $\nu$ and $\dot{\nu}$ only. They defined their metric, $\sigma^2_{\rm{TN}}$, as
\begin{equation}
\label{sigmatn}
    \sigma^2_{\rm{TN}} = \sigma^2_{\rm R} - \sigma^2_{\rm W}
\end{equation}
where $\sigma_{\rm R}$ is the measured rms of the residuals and $\sigma_{\rm W}$ is the typical ToA error for a given pulsar.
\citet{sc10} used $\sigma_{\rm{TN}}$, measured for a large number of pulsars, to derive a generic measure, $\sigma_{\rm M}$, for the timing noise based on the spin parameters.
\begin{equation}
\label{sigmam}
    \sigma_{\rm M} = K\,\,\nu^{a}\,\,\,|\dot{\nu}|^{b}\,\,\,T^{c}
\end{equation}
They used a large compilation of timing noise values from the literature and found that $K=7\pm4$, $a=-0.9\pm0.2$, $b=1.00\pm0.05$ and $c=1.9\pm0.2$ provided the best fit for the `normal' pulsars, i.e. those which are neither magnetars nor millisecond pulsars. They also recognised that there was a large scatter in $\sigma_{\rm M}$ for pulsars with similar spin parameters.
Finally, \citet{psj+19} have examined the timing noise of young pulsars using a Bayesian approach which fits simultaneously for the timing parameters and a power-law representation for the spectrum of timing noise. They find a mean value of $-5.2$ for the spectral index. Overall, their results are generally in line with those found by \citet{sc10}.

In this paper we examine the timing noise in a group of 129 mostly middle-aged pulsars (i.e. ages near 1~Myr) observed with the Parkes radio telescope on a regular basis since 2014. In Section~2 we briefly describe the observations and the data analysis, Section~3 presents the results, the implications of which are discussed in Section~4.

\section{Observations and data analysis}
Starting in 2007, under the auspices of Parkes project P574, we have observed non-millisecond pulsars with a monthly cadence. Prior to 2014 the sample largely contained young pulsars with a spin-down energy, $\dot{E}$ greater than $10^{35.5}$~ergs$^{-1}$ \citep{sgc+08}. Since 2014 January, the sample was augmented by a sample of bright, low $\dot{E}$ pulsars and it is this sample that we discuss here. Table~1 shows the basic parameters for the 129 pulsars in our sample. The table lists $\nu$ and $\dot{\nu}$, the spin-down energy, $\dot{E} = -4\pi^2I\nu\dot{\nu}$, the characteristic age, $\tau_c = -\nu/(2\dot{\nu})$, and the magnetic field strength, $B=3.2\times10^{19}\nu^{-3/2}\dot{\nu}^{1/2}$.

The data are taken and analysed using the methods extensively described in \citet{wjm+10} and \citet{jk18}. In brief, we use the Parkes radio telescope to observe at a centre frequency of 1369~MHz with a total bandwidth of 256~MHz subdivided into 1024 frequency channels. Observations are folded at the topocentric period of the pulsar and dedispersed then accumulated into 30~s of data termed subintegrations. Typical observations are 180~s in duration (i.e. 6 subintegrations). Observations of a calibrator signal and of the sky-calibrator Hydra~A allow polarization and flux density calibration.

\begin{figure}
    \centering
    \includegraphics[width=0.45\textwidth]{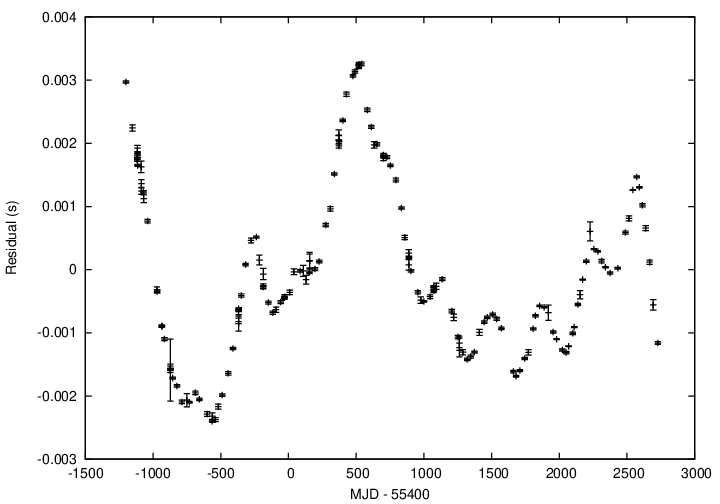}
    \caption{Post-fit residuals for 11 years of timing of PSR~J1705$-$1906 after fitting for $\nu$ and $\dot{\nu}$.}
    \label{fig:1705}
\end{figure}
For the purposes of this work, the data are excised of interference, calibrated, and then summed in both frequency in time to provide a single high-fidelity pulse profile per observation. Using standard techniques made available via the package {\sc psrchive} \citep{hvm04}, the profile is then compared (in the frequency domain) with a high signal-to-noise template to produce a ToA and associated error bar. The ensemble of ToAs for a given pulsar are then compared against a timing model using the package {\sc tempo2} \citep{hem06}. We initially derive a best-fit model using only $\nu$ and $\dot{\nu}$ and compute the rms of the residuals, $\sigma_{\rm R}$ for use in Equation~\ref{sigmatn}. We then fit for $\ddot{\nu}$ and its error, necessary for the computations of Equations~\ref{delta8}, \ref{sigmaz} and \ref{bi}. We use the $F$-test to determine whether the fitting of an extra parameter is justified. We find that, where {\sc tempo2} returns a significant value of $\ddot{\nu}$, in all but a handful of cases the $F$-test is passed at the 95\% (2-$\sigma$) confidence level. We therefore retain all the {\sc tempo2} values, noting that some false positives may occur but this does not affect our overall conclusions.

\section{Results}
Table~2 shows the results of the fitting. The table gives the time span of the observation (in years) followed by the median ToA error, $\sigma_{\rm W}$, in $\mu$s. $\sigma_{\rm TN}$ is given in column 4; for pulsars with $\sigma_R<2\sigma_{\rm W}$ we set an upper limit to $\sigma_{\rm TN}$ of $2\sigma_{\rm W}$ (see equation~\ref{sigmatn}). The sign of $\ddot{\nu}$ is listed along with the logarithm of its value. For pulsars for which the significance of $\ddot{\nu}$ is less then 2-$\sigma$, according to {\sc tempo2}, we set an upper limit equal to 2-$\sigma$. Otherwise we quote the 1-$\sigma$ error bar as reported by {\sc tempo2} converted into log. The final four columns list $n$, $\sigma_z$, $\Delta_8$ and $\sigma_{\rm M}$. The error bars on $\sigma_z$, $\Delta_8$ and $n$ are entirely dominated by the error in $\ddot{\nu}$ and therefore have the same value in the log as the error on $\ddot{\nu}$ as they all depend linearly on $\ddot{\nu}$.

Although the majority of pulsars have been observed for 4~yr, for 16 of pulsars the dataset extends up to 15~yr. Table~2 simply shows the values of $\sigma_z$ and $\Delta_8$ as given by equations~\ref{sigmaz} and \ref{delta8} with the appropriate values of $T$. However, for these 16 pulsars, we subdivided the data into 4~yr blocks (with partial overlap) and computed $\sigma_{\rm TN}$ and $\ddot{\nu}$ in each block. We then recomputed $\sigma_z$, $\Delta_8$ and $n$ using the average of $\ddot{\nu}$ and with $T=4$~yr. Results are shown in Table~3 which has a similar format to Table~2.

\begin{table*}
  \centering
  \caption{Results showing observational time-span ($T$), median ToA error ($\sigma_{\rm W}$), the sign and magnitude of $\ddot{\nu}$ and the metrics $n$, $\sigma_z$, $\Delta_8$, $\sigma_{\rm TN}$ and $\sigma_{\rm M}$.}
  \label{results}
\renewcommand{\arraystretch}{1.42}
\resizebox{\textwidth}{!}{
    \begin{tabular}{lccccccccc}
	Name & $T$ & $\sigma_{\rm W}$ & sign of $\ddot{\nu}$ & log$|\ddot{\nu}|$ & log$|n|$ & log($\sigma_z$) & log($\Delta_8$) & log($\sigma_{\rm TN}$) & log($\sigma_{\rm M}$)\\ 
	& (yr) & ($\mu$s) && (/s$^{-3}$) &&&& (/$\mu$s)& (/$\mu$s) \\ \hline
	J0034  $  -  $  0721  &  14.0  &  230  &   $   $   & $  <-26.73^{ }_{ }$   & $  <3.97^{  }_{}$   & $  <-10.09^{ }_{ }$   & $   <-1.60^{ }_{ }$   &  3.33  &  2.69
  \\ 
J0051  $  +  $  0423  &  4.1  &  1153  &   $   $   & $  <-24.70^{  }_{  }$   & $  <8.38^{  }_{ }$   & $  <-9.14^{ }_{ }$   & $   <-1.60^{ }_{ }$   &  <3.36  &  0.30
  \\ 
J0108  $  -  $  1431  &  15.0  &  466  &   $  +  $   & $  -26.40^{ + 0.04}_{ - 0.05}$   & $  5.54^{ + 0.04}_{- 0.05}$   & $  -9.71^{+ 0.04}_{- 0.05}$   & $   -1.25^{+ 0.04}_{- 0.05}$   &  3.14  &  2.09
  \\ 
J0134  $  -  $  2937  &  11.3  &  9  &   $  +  $   & $  -26.81^{ + 0.03}_{ - 0.03}$   & $  2.82^{ + 0.03}_{- 0.03}$   & $  -10.35^{+ 0.03}_{- 0.03}$   & $   -2.79^{+ 0.03}_{- 0.03}$   &  1.26  &  2.71
  \\ 
J0152  $  -  $  1637  &  13.1  &  87  &   $  +  $   & $  -27.38^{ + 0.10}_{ - 0.13}$   & $  2.15^{ + 0.10}_{- 0.13}$   & $  -10.80^{+ 0.10}_{- 0.13}$   & $   -2.39^{+ 0.10}_{- 0.13}$   &  <2.24  &  3.19
  \\ 
J0206  $  -  $  4028  &  11.1  &  165  &   $  +  $   & $  -25.70^{ + 0.02}_{ - 0.02}$   & $  3.54^{ + 0.02}_{- 0.02}$   & $  -9.26^{+ 0.02}_{- 0.02}$   & $   -1.04^{+ 0.02}_{- 0.02}$   &  3.18  &  3.16
  \\ 
J0255  $  -  $  5304  &  3.9  &  16  &   $   $   & $  <-25.72^{  }_{  }$   & $  <6.26^{  }_{ }$   & $  <-10.20^{ }_{ }$   & $   <-2.59^{ }_{ }$   &  2.18  &  0.85
  \\ 
J0343  $  -  $  3000  &  4.1  &  1548  &   $   $   & $  <-24.94^{  }_{  }$   & $  <8.46^{  }_{ }$   & $  <-9.38^{ }_{ }$   & $   <-0.98^{ }_{ }$   &  <3.49  &  0.49
  \\ 
J0401  $  -  $  7608  &  14.7  &  155  &   $   $   & $  -25.36^{ + 0.02}_{ - 0.02}$   & $  3.48^{ + 0.02}_{- 0.02}$   & $  -8.67^{+ 0.02}_{- 0.02}$   & $   -0.40^{+ 0.02}_{- 0.02}$   &  3.70  &  3.56
  \\ 
J0448  $  -  $  2749  &  4.1  &  115  &   $   $   & $  <-25.61^{  }_{  }$   & $  <5.00^{  }_{}$   & $  <-10.05^{ }_{ }$   & $   <-2.42^{ }_{}$   &  <2.36  &  1.58
  \\ 
J0452  $  -  $  1759  &  12.3  &  33  &   $  -  $   & $  -25.58^{ + 0.04}_{ - 0.05}$   & $  2.12^{ + 0.04}_{- 0.05}$   & $  -9.05^{+ 0.04}_{- 0.05}$   & $   -0.86^{+ 0.04}_{- 0.05}$   &  3.62  &  3.98
  \\ 
J0525  $  +  $  1115  &  3.9  &  71  &   $   $   & $  <-25.94^{  }_{  }$   & $  <4.97^{  }_{}$   & $  <-10.42^{ }_{ }$   & $   <-2.92^{ }_{ }$   &  <2.15  &  1.35
  \\ 
J0536  $  -  $  7543  &  14.0  &  141  &   $  -  $   & $  -27.27^{ + 0.15}_{ - 0.24}$   & $  3.49^{ + 0.15}_{- 0.24}$   & $  -10.63^{+ 0.15}_{- 0.24}$   & $   -2.02^{+ 0.15}_{- 0.24}$   &  2.69  &  2.70
  \\ 
J0601  $  -  $  0527  &  3.9  &  47  &   $  -  $   & $  -25.61^{ + 0.08}_{ - 0.09}$   & $  2.96^{ + 0.08}_{- 0.09}$   & $  -10.09^{+ 0.08}_{- 0.09}$   & $   -2.54^{+ 0.08}_{- 0.09}$   &  <1.97  &  2.54
  \\ 
J0624  $  -  $  0424  &  3.9  &  108  &   $  -  $   & $  -26.23^{ + 0.14}_{ - 0.23}$   & $  3.98^{ + 0.14}_{- 0.23}$   & $  -10.72^{+ 0.14}_{- 0.23}$   & $   -2.74^{+ 0.14}_{- 0.23}$   &  <2.33  &  1.88
  \\ 
J0630  $  -  $  2834  &  13.2  &  148  &   $  +  $   & $  -25.62^{ + 0.02}_{ - 0.03}$   & $  2.95^{ + 0.02}_{- 0.03}$   & $  -9.04^{+ 0.02}_{- 0.03}$   & $   -0.45^{+ 0.02}_{- 0.03}$   &  3.93  &  3.75
  \\ 
J0729  $  -  $  1836  &  11.3  &  82  &   $  +  $   & $  -24.00^{ + 0.02}_{ - 0.02}$   & $  2.56^{ + 0.02}_{- 0.02}$   & $  -7.55^{+ 0.02}_{- 0.02}$   & $   0.58^{+ 0.02}_{- 0.02}$   &  5.06  &  4.47
  \\ 
J0738  $  -  $  4042  &  9.9  &  26  &   $  +  $   & $  -24.41^{ + 0.02}_{ - 0.02}$   & $  4.04^{ + 0.02}_{- 0.02}$   & $  -8.07^{+ 0.02}_{- 0.02}$   & $   -0.13^{+ 0.02}_{- 0.02}$   &  4.15  &  3.36
  \\ 
J0809  $  -  $  4753  &  4.1  &  66  &   $  +  $   & $  -25.16^{ + 0.10}_{ - 0.14}$   & $  3.07^{ + 0.10}_{- 0.14}$   & $  -9.58^{+ 0.10}_{- 0.14}$   & $   -1.86^{+ 0.10}_{- 0.14}$   &  2.75  &  2.82
  \\ 
J0820  $  -  $  1350  &  3.7  &  42  &   $   $   & $  <-26.10^{  }_{  }$   & $  <3.53^{  }_{ }$   & $  <-10.62^{ }_{ }$   & $   <-2.59^{ }_{ }$   &  2.19  &  2.16
  \\ 
J0820  $  -  $  4114  &  4.1  &  598  &   $   $   & $  <-25.03^{  }_{  }$   & $  <7.74^{  }_{ }$   & $  <-9.45^{ }_{ }$   & $   <-1.72^{ }_{ }$   &  <3.08  &  0.55
  \\ 
J0837  $  -  $  4135  &  4.0  &  7  &   $  +  $   & $  -25.07^{ + 0.03}_{ - 0.03}$   & $  3.46^{ + 0.03}_{- 0.03}$   & $  -9.51^{+ 0.03}_{- 0.03}$   & $   -1.65^{+ 0.03}_{- 0.03}$   &  2.62  &  2.71
  \\ 
J0904  $  -  $  7459  &  3.8  &  298  &   $   $   & $  <-25.44^{  }_{  }$   & $  <4.45^{  }_{ }$   & $  <-9.93^{ }_{ }$   & $   <-2.24^{}_{ }$   &  <2.78  &  1.92
  \\ 
J0907  $  -  $  5157  &  4.0  &  16  &   $  +  $   & $  -24.61^{ + 0.03}_{ - 0.04}$   & $  3.07^{ + 0.03}_{- 0.04}$   & $  -9.05^{+ 0.03}_{- 0.04}$   & $   -1.68^{+ 0.03}_{- 0.04}$   &  2.67  &  2.94
  \\ 
J0924  $  -  $  5814  &  4.0  &  226  &   $  -  $   & $  -24.26^{ + 0.02}_{ - 0.02}$   & $  4.01^{ + 0.02}_{- 0.02}$   & $  -8.69^{+ 0.02}_{- 0.02}$   & $   -0.85^{+ 0.02}_{- 0.02}$   &  3.44  &  2.83
  \\ 
J0942  $  -  $  5552  &  3.9  &  37  &   $  +  $   & $  -22.92^{ + 0.01}_{ - 0.01}$   & $  3.66^{ + 0.01}_{- 0.01}$   & $  -7.40^{+ 0.01}_{- 0.01}$   & $   0.39^{+ 0.01}_{- 0.01}$   &  4.66  &  3.62
  \\ 
J1001  $  -  $  5507  &  4.0  &  19  &   $  +  $   & $  -24.20^{ + 0.06}_{ - 0.07}$   & $  2.83^{ + 0.06}_{- 0.07}$   & $  -8.64^{+ 0.06}_{- 0.07}$   & $   -0.50^{+ 0.06}_{- 0.07}$   &  3.92  &  3.57
  \\ 
J1034  $  -  $  3224  &  4.0  &  599  &   $   $   & $  <-25.34^{  }_{  }$   & $  <6.11^{  }_{ }$   & $  <-9.78^{ }_{ }$   & $   <-1.74^{ }_{ }$   &  <3.08  &  1.32
  \\ 
J1038  $  -  $  5831  &  3.9  &  113  &   $  -  $   & $  -24.81^{ + 0.03}_{ - 0.02}$   & $  4.50^{ + 0.03}_{- 0.02}$   & $  -9.29^{+ 0.03}_{- 0.02}$   & $   -1.51^{+ 0.03}_{- 0.02}$   &  2.73  &  2.26
  \\ 
J1047  $  -  $  6709  &  4.0  &  16  &   $  +  $   & $  -26.16^{ + 0.04}_{ - 0.04}$   & $  1.28^{ + 0.04}_{- 0.04}$   & $  -10.59^{+ 0.04}_{- 0.04}$   & $   -3.32^{+ 0.04}_{- 0.04}$   &  2.01  &  3.02
  \\ 
J1055  $  -  $  6022  &  5.7  &  1378  &   $  +  $   & $  -24.63^{ + 0.02}_{ - 0.02}$   & $  1.37^{ + 0.02}_{- 0.02}$   & $  -8.76^{+ 0.02}_{- 0.02}$   & $   -0.66^{+ 0.02}_{- 0.02}$   &  3.60  &  4.30
  \\ 
J1056  $  -  $  6258  &  4.0  &  22  &   $  +  $   & $  -24.79^{ + 0.14}_{ - 0.20}$   & $  2.98^{ + 0.14}_{- 0.20}$   & $  -9.23^{+ 0.14}_{- 0.20}$   & $   -1.62^{+ 0.14}_{- 0.20}$   &  3.02  &  2.99
  \\ 
J1136  $  -  $  5525  &  4.0  &  42  &   $  -  $   & $  -23.60^{ + 0.05}_{ - 0.06}$   & $  3.29^{ + 0.05}_{- 0.06}$   & $  -8.04^{+ 0.05}_{- 0.06}$   & $   -0.50^{+ 0.05}_{- 0.06}$   &  3.88  &  3.40
  \\ 
J1146  $  -  $  6030  &  4.0  &  24  &   $   $   & $  <-25.73^{  }_{  }$   & $  <2.08^{  }_{ }$   & $  <-10.17^{ }_{ }$   & $   <-2.75^{ }_{ }$   &  1.99  &  2.89
  \\ 
J1157  $  -  $  6224  &  4.0  &  28  &   $  +  $   & $  -25.19^{ + 0.14}_{ - 0.20}$   & $  2.43^{ + 0.14}_{- 0.20}$   & $  -9.63^{+ 0.14}_{- 0.20}$   & $   -2.05^{+ 0.14}_{- 0.20}$   &  2.61  &  3.05
  \\ 
J1243  $  -  $  6423  &  4.0  &  3  &   $  +  $   & $  -24.60^{ + 0.04}_{ - 0.05}$   & $  2.85^{ + 0.04}_{- 0.05}$   & $  -9.04^{+ 0.04}_{- 0.05}$   & $   -1.47^{+ 0.04}_{- 0.05}$   &  2.87  &  3.13
  \\ 
J1306  $  -  $  6617  &  3.8  &  11  &   $  +  $   & $  -23.84^{ + 0.01}_{ - 0.01}$   & $  3.59^{ + 0.01}_{- 0.01}$   & $  -8.34^{+ 0.01}_{- 0.01}$   & $   -0.71^{+ 0.01}_{- 0.01}$   &  3.58  &  3.12
  \\ 
J1317  $  -  $  6302  &  4.0  &  144  &   $  +  $   & $  -25.19^{ + 0.15}_{ - 0.23}$   & $  5.05^{ + 0.15}_{- 0.23}$   & $  -9.63^{+ 0.15}_{- 0.23}$   & $   -2.24^{+ 0.15}_{- 0.23}$   &  <2.46  &  1.66  \\ 
\\[+1em]
	\renewcommand{\tabcolsep}{40mm}
    \end{tabular}}
\end{table*}
\begin{table*}
\addtocounter{table}{-1}
    \centering
    \caption{Results (continued)}
\renewcommand{\arraystretch}{1.42}
\resizebox{\textwidth}{!}{
    \begin{tabular}{lccccccccc}
	Name & $T$ & $\sigma_{\rm W}$ & sign of $\ddot{\nu}$ & log$|\ddot{\nu}|$ & log$|n|$ & log($\sigma_z$) & log($\Delta_8$) & log($\sigma_{\rm TN}$) & log($\sigma_{\rm M}$)\\ 
	& (yr) & ($\mu$s) && (/s$^{-3}$) &&&& (/$\mu$s)& (/$\mu$s) \\ \hline
	J1319  $  -  $  6056  &  4.0  &  60  &   $  +  $   & $  -25.22^{ + 0.04}_{ - 0.06}$   & $  2.77^{ + 0.04}_{- 0.06}$   & $  -9.67^{+ 0.04}_{- 0.06}$   & $   -2.24^{+ 0.04}_{- 0.06}$   &  2.06  &  2.80
  \\ 
J1326  $  -  $  5859  &  4.0  &  7  &   $  +  $   & $  -23.98^{ + 0.02}_{ - 0.04}$   & $  3.98^{ + 0.02}_{- 0.04}$   & $  -8.43^{+ 0.02}_{- 0.04}$   & $   -0.77^{+ 0.02}_{- 0.04}$   &  3.54  &  2.91
  \\ 
J1326  $  -  $  6408  &  4.0  &  91  &   $  -  $   & $  -25.90^{ + 0.16}_{ - 0.27}$   & $  2.82^{ + 0.16}_{- 0.27}$   & $  -10.35^{+ 0.16}_{- 0.27}$   & $   -2.47^{+ 0.16}_{- 0.27}$   &  2.23  &  2.62
  \\ 
J1326  $  -  $  6700  &  4.0  &  54  &   $  +  $   & $  -24.97^{ + 0.15}_{ - 0.25}$   & $  2.77^{ + 0.15}_{- 0.25}$   & $  -9.42^{+ 0.15}_{- 0.25}$   & $   -1.71^{+ 0.15}_{- 0.25}$   &  3.51  &  3.04
  \\ 
J1327  $  -  $  6222  &  4.0  &  17  &   $  +  $   & $  -23.36^{ + 0.02}_{ - 0.02}$   & $  3.22^{ + 0.02}_{- 0.02}$   & $  -7.80^{+ 0.02}_{- 0.02}$   & $   -0.10^{+ 0.02}_{- 0.02}$   &  4.10  &  3.62
  \\ 
J1327  $  -  $  6301  &  4.0  &  41  &   $  +  $   & $  -25.37^{ + 0.10}_{ - 0.13}$   & $  2.14^{ + 0.10}_{- 0.13}$   & $  -9.82^{+ 0.10}_{- 0.13}$   & $   -2.55^{+ 0.10}_{- 0.13}$   &  <1.92  &  2.98
  \\ 
J1328  $  -  $  4357  &  4.0  &  46  &   $  -  $   & $  -24.43^{ + 0.05}_{ - 0.06}$   & $  3.81^{ + 0.05}_{- 0.06}$   & $  -8.88^{+ 0.05}_{- 0.06}$   & $   -1.18^{+ 0.05}_{- 0.06}$   &  3.07  &  2.78
  \\ 
J1338  $  -  $  6204  &  4.0  &  347  &   $   $   & $  <-25.37^{  }_{  }$   & $  <2.63^{  }_{ }$   & $  <-9.81^{ }_{ }$   & $   <-1.75^{ }_{ }$   &  <2.84  &  3.05
  \\ 
J1340  $  -  $  6456  &  4.0  &  261  &   $  +  $   & $  -24.28^{ + 0.04}_{ - 0.05}$   & $  3.03^{ + 0.04}_{- 0.05}$   & $  -8.73^{+ 0.04}_{- 0.05}$   & $   -1.17^{+ 0.04}_{- 0.05}$   &  3.09  &  3.19
  \\ 
J1352  $  -  $  6803  &  4.0  &  454  &   $  +  $   & $  -25.42^{ + 0.16}_{ - 0.28}$   & $  3.78^{ + 0.16}_{- 0.28}$   & $  -9.87^{+ 0.16}_{- 0.28}$   & $   -2.09^{+ 0.16}_{- 0.28}$   &  <2.96  &  2.33
  \\ 
J1356  $  -  $  5521  &  4.0  &  362  &   $   $   & $  <-25.41^{  }_{  }$   & $  <3.97^{  }_{ }$   & $  <-9.85^{ }_{ }$   & $   <-2.17^{ }_{ }$   &  <2.86  &  2.21
  \\ 
J1357  $  -  $  62  &  4.0  &  36  &   $  -  $   & $  -24.69^{ + 0.02}_{ - 0.02}$   & $  1.96^{ + 0.02}_{- 0.02}$   & $  -9.13^{+ 0.02}_{- 0.02}$   & $   -1.50^{+ 0.02}_{- 0.02}$   &  2.78  &  3.56
  \\ 
J1401  $  -  $  6357  &  4.0  &  20  &   $  -  $   & $  -23.63^{ + 0.03}_{ - 0.04}$   & $  3.77^{ + 0.03}_{- 0.04}$   & $  -8.08^{+ 0.03}_{- 0.04}$   & $   -0.18^{+ 0.03}_{- 0.04}$   &  4.14  &  3.28
  \\ 
J1418  $  -  $  3921  &  4.0  &  469  &   $   $   & $  <-25.38^{  }_{  }$   & $  <4.84^{  }_{ }$   & $  <-9.82^{ }_{ }$   & $   <-1.81^{ }_{}$   &  <2.97  &  1.92
  \\
J1424  $  -  $  5822  &  4.0  &  109  &   $   $   & $  <-24.56^{  }_{  }$   & $  <2.94^{  }_{ }$   & $  <-9.00^{ }_{ }$   & $   <-1.46^{ }_{ }$   &  3.32  &  3.09
  \\ 
J1428  $  -  $  5530  &  4.0  &  42  &   $  +  $   & $  -25.28^{ + 0.10}_{ - 0.13}$   & $  3.34^{ + 0.10}_{- 0.13}$   & $  -9.72^{+ 0.10}_{- 0.13}$   & $   -1.99^{+ 0.10}_{- 0.13}$   &  2.48  &  2.61
  \\ 
J1430  $  -  $  6623  &  4.0  &  11  &   $  +  $   & $  -25.99^{ + 0.10}_{ - 0.12}$   & $  2.81^{ + 0.10}_{- 0.12}$   & $  -10.43^{+ 0.10}_{- 0.12}$   & $   -2.56^{+ 0.10}_{- 0.12}$   &  1.92  &  2.58
  \\ 
J1435  $  -  $  5954  &  4.0  &  259  &   $   $   & $  <-25.38^{  }_{ }$   & $  <3.26^{ }_{ }$   & $  <-9.83^{ }_{ }$   & $   <-2.18^{ }_{ }$   &  <2.71  &  2.56
  \\ 
J1456  $  -  $  6843  &  10.1  &  32  &   $  +  $   & $  -26.57^{ + 0.09}_{ - 0.11}$   & $  3.70^{ + 0.09}_{- 0.11}$   & $  -10.21^{+ 0.09}_{- 0.11}$   & $   -2.41^{+ 0.09}_{- 0.11}$   &  2.17  &  2.41
  \\ 
J1522  $  -  $  5829  &  4.0  &  51  &   $  -  $   & $  -24.44^{ + 0.01}_{ - 0.01}$   & $  3.74^{ + 0.01}_{- 0.01}$   & $  -8.89^{+ 0.01}_{- 0.01}$   & $   -1.31^{+ 0.01}_{- 0.01}$   &  2.94  &  2.76
  \\ 
J1534  $  -  $  5334  &  4.0  &  48  &   $   $   & $  <-26.27^{  }_{  }$   & $  <3.83^{  }_{ }$   & $  <-10.72^{ }_{ }$   & $   <-2.61^{ }_{ }$   &  2.13  &  2.02
  \\ 
J1534  $  -  $  5405  &  4.0  &  91  &   $  -  $   & $  -23.29^{ + 0.01}_{ - 0.02}$   & $  4.72^{ + 0.01}_{- 0.02}$   & $  -7.74^{+ 0.01}_{- 0.02}$   & $   -0.30^{+ 0.01}_{- 0.02}$   &  3.46  &  2.79
  \\ 
J1536  $  -  $  5433  &  4.0  &  1000  &   $  -  $   & $  -24.08^{ + 0.04}_{ - 0.04}$   & $  5.27^{ + 0.04}_{- 0.04}$   & $  -8.52^{+ 0.04}_{- 0.04}$   & $   -0.61^{+ 0.04}_{- 0.04}$   &  3.73  &  2.32
  \\ 
J1544  $  -  $  5308  &  4.0  &  19  &   $  +  $   & $  -25.74^{ + 0.10}_{ - 0.12}$   & $  4.45^{ + 0.10}_{- 0.12}$   & $  -10.18^{+ 0.10}_{- 0.12}$   & $   -2.96^{+ 0.10}_{- 0.12}$   &  1.63  &  1.62
  \\ 
J1555  $  -  $  3134  &  4.0  &  117  &   $   $   & $  <-25.97^{  }_{  }$   & $  <5.59^{  }_{ }$   & $  <-10.41^{ }_{ }$   & $   <-2.72^{ }_{ }$   &  <2.37  &  1.12
  \\ 
J1557  $  -  $  4258  &  4.0  &  28  &   $  -  $   & $  -25.19^{ + 0.02}_{ - 0.03}$   & $  4.34^{ + 0.02}_{- 0.03}$   & $  -9.64^{+ 0.02}_{- 0.03}$   & $   -2.15^{+ 0.02}_{- 0.03}$   &  2.09  &  2.06
  \\ 
J1559  $  -  $  4438  &  4.0  &  6  &   $   $   & $  <-25.55^{  }_{ }$   & $  <2.66^{  }_{ }$   & $  <-10.00^{ }_{ }$   & $   -2.62^{ }_{ }$   &  2.14  &  2.67
  \\ 
J1604  $  -  $  4909  &  4.0  &  16  &   $  -  $   & $  -23.92^{ + 0.01}_{ - 0.01}$   & $  4.72^{ + 0.01}_{- 0.01}$   & $  -8.36^{+ 0.01}_{- 0.01}$   & $   -0.87^{+ 0.01}_{- 0.01}$   &  3.41  &  2.50
  \\ 
J1605  $  -  $  5257  &  4.0  &  100  &   $   $   & $  <-25.73^{  }_{  }$   & $  <4.91^{  }_{ }$   & $  <-10.18^{ }_{ }$   & $   <-2.38^{ }_{ }$   &  2.32  &  1.62
  \\
J1613  $  -  $  4717  &  4.0  &  86  &   $   $   & $  <-25.92^{  }_{  }$   & $  <3.23^{  }_{ }$   & $  <-10.36^{}_{ }$   & $   <-2.81^{ }_{ }$   &  <2.24  &  2.28
  \\ 
J1623  $  -  $  4256  &  4.0  &  127  &   $  -  $   & $  -24.30^{ + 0.03}_{ - 0.04}$   & $  4.45^{ + 0.03}_{- 0.04}$   & $  -8.75^{+ 0.03}_{- 0.04}$   & $   -1.21^{+ 0.03}_{- 0.04}$   &  3.12  &  2.46
  \\ 
J1626  $  -  $  4537  &  4.0  &  228  &   $  +  $   & $  -25.02^{ + 0.11}_{ - 0.16}$   & $  1.84^{ + 0.11}_{- 0.16}$   & $  -9.47^{+ 0.11}_{- 0.16}$   & $   -1.93^{+ 0.11}_{- 0.16}$   &  2.63  &  3.41
  \\ 
J1630  $  -  $  4733  &  4.5  &  466  &   $  -  $   & $  -24.20^{ + 0.10}_{ - 0.12}$   & $  2.38^{ + 0.10}_{- 0.12}$   & $  -8.54^{+ 0.10}_{- 0.12}$   & $   -0.76^{+ 0.10}_{- 0.12}$   &  3.54  &  3.72
  \\ 
J1632  $  -  $  4621  &  4.0  &  276  &   $  +  $   & $  -24.69^{ + 0.02}_{ - 0.02}$   & $  2.25^{ + 0.02}_{- 0.02}$   & $  -9.13^{+ 0.02}_{- 0.02}$   & $   -0.93^{+ 0.02}_{- 0.02}$   &  3.33  &  3.64
  \\ 
J1633  $  -  $  4453  &  4.0  &  66  &   $  -  $   & $  -24.96^{ + 0.04}_{ - 0.04}$   & $  2.38^{ + 0.04}_{- 0.04}$   & $  -9.40^{+ 0.04}_{- 0.04}$   & $   -1.79^{+ 0.04}_{- 0.04}$   &  2.53  &  3.20
  \\ 
J1633  $  -  $  5015  &  4.0  &  20  &   $  +  $   & $  -25.62^{ + 0.08}_{ - 0.10}$   & $  1.86^{ + 0.08}_{- 0.10}$   & $  -10.06^{+ 0.08}_{- 0.10}$   & $   -2.54^{+ 0.08}_{- 0.10}$   &  1.93  &  3.09
  \\ 
J1646  $  -  $  6831  &  4.0  &  206  &   $   $   & $  <-26.38^{  }_{  }$   & $  <3.91^{  }_{ }$   & $  <-10.82^{ }_{ }$   & $   <-2.59^{ }_{ }$   &  2.63  &  1.97
  \\ 
J1649  $  -  $  3805  &  4.0  &  127  &   $   $   & $  <-25.61^{  }_{  }$   & $  <5.55^{  }_{ }$   & $  <-10.06^{ }_{ }$   & $   <-2.66^{ }_{ }$   &  <2.41  &  1.20  \\ 

    \end{tabular}}
\end{table*}
\begin{table*}
\addtocounter{table}{-1}
    \centering
    \caption{Results (continued)}
\renewcommand{\arraystretch}{1.42}
\resizebox{\textwidth}{!}{
    \begin{tabular}{lccccccccc}
	Name & $T$ & $\sigma_{\rm W}$ & sign of $\ddot{\nu}$ & log$|\ddot{\nu}|$ & log$|n|$ & log($\sigma_z$) & log($\Delta_8$) & log($\sigma_{\rm TN}$) & log($\sigma_{\rm M}$)\\ 
	& (yr) & ($\mu$s) && (/s$^{-3}$) &&&& (/$\mu$s)& (/$\mu$s) \\ \hline
	J1651  $  -  $  4246  &  3.9  &  149  &   $  -  $   & $  -24.18^{ + 0.04}_{ - 0.04}$   & $  4.33^{ + 0.04}_{- 0.04}$   & $  -8.66^{+ 0.04}_{- 0.04}$   & $   -0.78^{+ 0.04}_{- 0.04}$   &  3.37  &  2.70
  \\ 
J1651  $  -  $  5222  &  4.0  &  53  &   $  +  $   & $  -25.37^{ + 0.05}_{ - 0.05}$   & $  3.51^{ + 0.05}_{- 0.05}$   & $  -9.81^{+ 0.05}_{- 0.05}$   & $   -2.04^{+ 0.05}_{- 0.05}$   &  2.30  &  2.50
  \\ 
J1652  $  -  $  2404  &  4.0  &  133  &   $   $   & $  <-26.05^{  }_{  }$   & $  <3.65^{  }_{ }$   & $  <-10.49^{ }_{ }$   & $   <-2.28^{ }_{ }$   &  2.50  &  2.26
  \\ 
J1653  $  -  $  3838  &  4.0  &  30  &   $  -  $   & $  -24.26^{ + 0.04}_{ - 0.04}$   & $  3.32^{ + 0.04}_{- 0.04}$   & $  -8.70^{+ 0.04}_{- 0.04}$   & $   -1.24^{+ 0.04}_{- 0.04}$   &  3.09  &  3.02
  \\ 
J1653  $  -  $  4249  &  4.0  &  182  &   $  +  $   & $  -24.44^{ + 0.09}_{ - 0.11}$   & $  3.56^{ + 0.09}_{- 0.11}$   & $  -8.88^{+ 0.09}_{- 0.11}$   & $   -1.12^{+ 0.09}_{- 0.11}$   &  3.39  &  2.93
  \\ 
J1700  $  -  $  3312  &  4.0  &  320  &   $  -  $   & $  -25.23^{ + 0.14}_{ - 0.19}$   & $  3.84^{ + 0.14}_{- 0.19}$   & $  -9.67^{+ 0.14}_{- 0.19}$   & $   -1.57^{+ 0.14}_{- 0.19}$   &  3.02  &  2.54
  \\ 
J1701  $  -  $  3726  &  4.1  &  183  &   $   $   & $  <-26.76^{  }_{ }$   & $  <2.31^{  }_{ }$   & $  <-11.19^{ }_{ }$   & $   <-2.82^{ }_{ }$   &  <2.56  &  2.65
  \\ 
J1703  $  -  $  3241  &  4.0  &  61  &   $  +  $   & $  -25.90^{ + 0.12}_{ - 0.17}$   & $  4.69^{ + 0.12}_{- 0.17}$   & $  -10.35^{+ 0.12}_{- 0.17}$   & $   -2.29^{+ 0.12}_{- 0.17}$   &  2.23  &  1.75
  \\
J1703  $  -  $  4851  &  4.0  &  445  &   $   $   & $  <-25.32^{  }_{  }$   & $  <3.72^{  }_{ }$   & $  <-9.76^{ }_{ }$   & $   <-1.64^{ }_{ }$   &  3.13  &  2.56
  \\ 
J1705  $  -  $  1906  &  11.4  &  26  &   $   $   & $  <-26.18^{  }_{  }$   & $  <1.02^{  }_{ }$   & $  <-9.71^{ }_{ }$   & $   <-1.81^{ }_{ }$   &  3.05  &  4.07
  \\ 
J1705  $  -  $  3423  &  4.0  &  53  &   $  -  $   & $  -24.30^{ + 0.07}_{ - 0.09}$   & $  3.86^{ + 0.07}_{- 0.09}$   & $  -8.75^{+ 0.07}_{- 0.09}$   & $   -1.37^{+ 0.07}_{- 0.09}$   &  3.10  &  2.70
  \\ 
J1707  $  -  $  4053  &  4.0  &  145  &   $  +  $   & $  -25.20^{ + 0.12}_{ - 0.15}$   & $  3.52^{ + 0.12}_{- 0.15}$   & $  -9.64^{+ 0.12}_{- 0.15}$   & $   -1.90^{+ 0.12}_{- 0.15}$   &  2.58  &  2.56
  \\ 
J1707  $  -  $  4729  &  4.0  &  91  &   $  -  $   & $  -25.59^{ + 0.14}_{ - 0.22}$   & $  2.30^{ + 0.14}_{- 0.22}$   & $  -10.04^{+ 0.14}_{- 0.22}$   & $   -2.64^{+ 0.14}_{- 0.22}$   &  <2.26  &  2.84
  \\ 
J1708  $  -  $  3426  &  4.0  &  113  &   $  -  $   & $  -25.54^{ + 0.12}_{ - 0.17}$   & $  2.73^{ + 0.12}_{- 0.17}$   & $  -9.99^{+ 0.12}_{- 0.17}$   & $   -2.17^{+ 0.12}_{- 0.17}$   &  2.40  &  2.81
  \\ 
J1709  $  -  $  1640  &  3.9  &  19  &   $  +  $   & $  -24.76^{ + 0.06}_{ - 0.08}$   & $  3.07^{ + 0.06}_{- 0.08}$   & $  -9.24^{+ 0.06}_{- 0.08}$   & $   -1.47^{+ 0.06}_{- 0.08}$   &  2.94  &  2.99
  \\ 
J1715  $  -  $  4034  &  4.0  &  552  &   $  +  $   & $  -25.45^{ + 0.14}_{ - 0.21}$   & $  4.51^{ + 0.14}_{- 0.21}$   & $  -9.90^{+ 0.14}_{- 0.21}$   & $   -1.61^{+ 0.14}_{- 0.21}$   &  3.01  &  2.16
  \\ 
J1717  $  -  $  3425  &  4.0  &  63  &   $  -  $   & $  -24.34^{ + 0.16}_{ - 0.24}$   & $  3.14^{ + 0.16}_{- 0.24}$   & $  -8.78^{+ 0.16}_{- 0.24}$   & $   -0.99^{+ 0.16}_{- 0.24}$   &  3.73  &  3.20
  \\ 
J1717  $  -  $  4054  &  11.4  &  100  &   $  +  $   & $  -24.95^{ + 0.05}_{ - 0.05}$   & $  3.76^{ + 0.05}_{- 0.05}$   & $  -8.49^{+ 0.05}_{- 0.05}$   & $   -0.12^{+ 0.05}_{- 0.05}$   &  4.37  &  3.50
  \\ 
J1717  $  -  $  5800  &  5.9  &  306  &   $   $   & $  <-26.63^{  }_{  }$   & $  <3.33^{  }_{ }$   & $  <-10.75^{ }_{ }$   & $   <-3.10^{ }_{ }$   &  <2.79  &  2.15
  \\ 
J1718  $  -  $  3718  &  4.8  &  10828  &   $  +  $   & $  -24.67^{ + 0.17}_{ - 0.26}$   & $  0.50^{ + 0.17}_{- 0.26}$   & $  -8.96^{+ 0.17}_{- 0.26}$   & $   -0.39^{+ 0.17}_{- 0.26}$   &  4.29  &  4.78
  \\ 
J1719  $  -  $  4006  &  4.0  &  99  &   $   $   & $  <-25.50^{  }_{  }$   & $  <1.88^{  }_{ }$   & $  <-9.95^{ }_{ }$   & $   <-2.70^{ }_{ }$   &  <2.29  &  3.03
  \\ 
J1720  $  -  $  2933  &  4.0  &  143  &   $  -  $   & $  -25.52^{ + 0.06}_{ - 0.06}$   & $  4.13^{ + 0.06}_{- 0.06}$   & $  -9.96^{+ 0.06}_{- 0.06}$   & $   -2.19^{+ 0.06}_{- 0.06}$   &  <2.46  &  2.11
  \\ 
J1722  $  -  $  3207  &  4.0  &  25  &   $  -  $   & $  -24.80^{ + 0.01}_{ - 0.02}$   & $  4.65^{ + 0.01}_{- 0.02}$   & $  -9.25^{+ 0.01}_{- 0.02}$   & $   -1.59^{+ 0.01}_{- 0.02}$   &  2.67  &  2.16
  \\ 
J1722  $  -  $  3632  &  4.0  &  97  &   $  -  $   & $  -25.51^{ + 0.12}_{ - 0.18}$   & $  1.99^{ + 0.12}_{- 0.18}$   & $  -9.96^{+ 0.12}_{- 0.18}$   & $   -2.38^{+ 0.12}_{- 0.18}$   &  <2.29  &  3.10
  \\ 
J1727  $  -  $  2739  &  4.1  &  319  &   $   $   & $  <-25.88^{  }_{  }$   & $  <4.39^{  }_{ }$   & $  <-10.31^{ }_{ }$   & $   <-2.22^{ }_{ }$   &  <2.81  &  1.94
  \\ 
J1733  $  -  $  2228  &  4.0  &  241  &   $   $   & $  <-25.70^{  }_{  }$   & $  <6.67^{  }_{ }$   & $  <-9.24^{ }_{ }$   & $   <-0.87^{ }_{ }$   &  <2.82  &  1.67
  \\ 
J1738  $  -  $  3211  &  4.0  &  116  &   $   $   & $  <-25.57^{  }_{  }$   & $  <4.30^{  }_{ }$   & $  <-10.01^{ }_{ }$   & $   <-2.15^{ }_{ }$   &  2.58  &  2.04
  \\ 
J1739  $  -  $  3131  &  4.0  &  119  &   $  -  $   & $  -23.02^{ + 0.02}_{ - 0.03}$   & $  3.71^{ + 0.02}_{- 0.03}$   & $  -7.47^{+ 0.02}_{- 0.03}$   & $   0.23^{+ 0.02}_{- 0.03}$   &  3.98  &  3.54
  \\ 
J1741  $  -  $  2733  &  4.0  &  428  &   $   $   & $  <-25.51^{  }_{ -}$   & $  <5.96^{  }_{ }$   & $  <-9.95^{ }_{ }$   & $   <-2.02^{ }_{ }$   &  <2.93  &  1.26
  \\ 
J1741  $  -  $  3016  &  4.0  &  521  &   $   $   & $  <-25.41^{  }_{  }$   & $  <3.52^{  }_{ }$   & $  <-9.85^{ }_{ }$   & $   <-1.60^{ }_{ }$   &  3.03  &  2.67
  \\ 
J1741  $  -  $  3927  &  4.0  &  40  &   $  +  $   & $  -23.48^{ + 0.001}_{ - 0.001}$   & $  4.98^{ + 0.001}_{- 0.001}$   & $  -7.92^{+ 0.001}_{- 0.001}$   & $   -0.24^{+ 0.001}_{- 0.001}$   &  4.05  &  2.67
  \\ 
J1743  $  -  $  3150  &  4.0  &  366  &   $   $   & $  <-25.72^{  }_{  }$   & $  <1.27^{  }_{ }$   & $  <-10.16^{ }_{ }$   & $   <-1.80^{ }_{ }$   &  2.92  &  3.68
  \\
J1749  $  -  $  3002  &  3.9  &  317  &   $  +  $   & $  -24.32^{ + 0.05}_{ - 0.05}$   & $  3.23^{ + 0.05}_{- 0.05}$   & $  -8.80^{+ 0.05}_{- 0.05}$   & $   -1.06^{+ 0.05}_{- 0.05}$   &  3.31  &  3.12
  \\ 
J1751  $  -  $  4657  &  4.0  &  31  &   $   $   & $  <-26.11^{  }_{  }$   & $  <3.27^{  }_{ }$   & $  <-10.56^{ }_{ }$   & $   <-2.71^{ }_{ }$   &  2.06  &  2.27
  \\
J1752  $  -  $  2806  &  3.9  &  8  &   $   $   & $  <-24.76^{  }_{  }$   & $  <2.67^{  }_{ }$   & $  <-9.24^{ }_{ }$   & $   <-1.53^{ }_{ }$   &  3.21  &  3.17
  \\ 
J1807  $  -  $  0847  &  3.9  &  8  &   $  +  $   & $  -25.86^{ + 0.07}_{ - 0.08}$   & $  4.86^{ + 0.07}_{- 0.08}$   & $  -10.34^{+ 0.07}_{- 0.08}$   & $   -3.17^{+ 0.07}_{- 0.08}$   &  1.25  &  1.31
  \\ 
J1816  $  -  $  2650  &  4.0  &  274  &   $  +  $   & $  -25.13^{ + 0.13}_{ - 0.17}$   & $  6.32^{ + 0.13}_{- 0.17}$   & $  -9.57^{+ 0.13}_{- 0.17}$   & $   -1.82^{+ 0.13}_{- 0.17}$   &  <2.74  &  1.20
  \\ 
J1817  $  -  $  3618  &  4.1  &  46  &   $  +  $   & $  -24.03^{ + 0.03}_{ - 0.03}$   & $  4.05^{ + 0.03}_{- 0.03}$   & $  -8.45^{+ 0.03}_{- 0.03}$   & $   -0.87^{+ 0.03}_{- 0.03}$   &  3.46  &  2.83  \\ 

    \end{tabular}}
\end{table*}
\begin{table*}
\addtocounter{table}{-1}
    \centering
    \caption{Results (continued)}
\renewcommand{\arraystretch}{1.42}
\resizebox{\textwidth}{!}{
    \begin{tabular}{lccccccccc}
	Name & $T$ & $\sigma_{\rm W}$ & sign of $\ddot{\nu}$ & log$|\ddot{\nu}|$ & log$|n|$ & log($\sigma_z$) & log($\Delta_8$) & log($\sigma_{\rm TN}$) & log($\sigma_{\rm M}$)\\ 
	& (yr) & ($\mu$s) && (/s$^{-3}$) &&&& (/$\mu$s)& (/$\mu$s) \\ \hline
	J1820  $  -  $  0427  &  4.2  &  20  &   $  -  $   & $  -24.40^{ + 0.01}_{ - 0.02}$   & $  3.34^{ + 0.01}_{- 0.02}$   & $  -8.80^{+ 0.01}_{- 0.02}$   & $   -1.03^{+ 0.01}_{- 0.02}$   &  3.23  &  3.10
  \\ 
J1822  $  -  $  2256  &  4.1  &  294  &   $   $   & $  <-25.70^{  }_{  }$   & $  <4.85^{  }_{ }$   & $  <-10.12^{ }_{ }$   & $   <-1.86^{ }_{ }$   &  2.81  &  1.87
  \\ 
J1822  $  -  $  4209  &  4.1  &  172  &   $  +  $   & $  -25.15^{ + 0.08}_{ - 0.09}$   & $  4.49^{ + 0.08}_{- 0.09}$   & $  -9.57^{+ 0.08}_{- 0.09}$   & $   -1.92^{+ 0.08}_{- 0.09}$   &  <2.54  &  2.08
  \\ 
J1823  $  -  $  3106  &  4.1  &  14  &   $  -  $   & $  -24.44^{ + 0.03}_{ - 0.03}$   & $  2.99^{ + 0.03}_{- 0.03}$   & $  -8.86^{+ 0.03}_{- 0.03}$   & $   -1.42^{+ 0.03}_{- 0.03}$   &  2.92  &  3.10
  \\ 
J1829  $  -  $  1751  &  3.9  &  11  &   $  -  $   & $  -23.71^{ + 0.06}_{ - 0.08}$   & $  3.28^{ + 0.06}_{- 0.08}$   & $  -8.17^{+ 0.06}_{- 0.08}$   & $   -0.72^{+ 0.06}_{- 0.08}$   &  3.74  &  3.30
  \\ 
J1845  $  -  $  0434  &  10.6  &  117  &   $  -  $   & $  -24.57^{ + 0.05}_{ - 0.05}$   & $  2.39^{ + 0.05}_{- 0.05}$   & $  -8.17^{+ 0.05}_{- 0.05}$   & $   -0.08^{+ 0.05}_{- 0.05}$   &  4.40  &  4.22
  \\ 
J1848  $  -  $  0123  &  4.0  &  33.9  &   $ + $   & $  -25.02^{ + 0.06 }_{ -0.07 }$   & $  3.00^{ + 0.06 }_{ - 0.07 }$   & $ -9.74^{ + 0.06 }_{ - 0.07 }$   & $   -1.68^{ + 0.06 }_{ -0.07 }$   &  2.30  &  2.93
  \\ 
J1852  $  -  $  0635  &  3.9  &  119  &   $  +  $   & $  -24.66^{ + 0.04}_{ - 0.06}$   & $  2.16^{ + 0.04}_{- 0.06}$   & $  -9.14^{+ 0.04}_{- 0.06}$   & $   -1.47^{+ 0.04}_{- 0.06}$   &  2.97  &  3.46
  \\ 
J1852  $  -  $  2610  &  4.1  &  153  &   $  +  $   & $  -25.44^{ + 0.12}_{ - 0.15}$   & $  5.24^{ + 0.12}_{- 0.15}$   & $  -9.86^{+ 0.12}_{- 0.15}$   & $   -2.34^{+ 0.12}_{- 0.15}$   &  <2.49  &  1.51
  \\ 
J1900  $  -  $  2600  &  13.2  &  55  &   $  -  $   & $  -26.23^{ + 0.01}_{ - 0.02}$   & $  4.51^{ + 0.01}_{- 0.02}$   & $  -9.65^{+ 0.01}_{- 0.02}$   & $   -1.37^{+ 0.01}_{- 0.02}$   &  3.00  &  2.54
  \\ 
J1913  $  -  $  0440  &  3.9  &  18  &   $  -  $   & $  -24.57^{ + 0.03}_{ - 0.03}$   & $  3.99^{ + 0.03}_{- 0.03}$   & $  -9.05^{+ 0.03}_{- 0.03}$   & $   -1.18^{+ 0.03}_{- 0.03}$   &  3.13  &  2.67
  \\ 
J1941  $  -  $  2602  &  4.1  &  29  &   $   $   & $  <-25.97^{  }_{  }$   & $  <2.88^{  }_{ }$   & $  <-10.39^{ }_{ }$   & $   <-2.80^{ }_{ }$   &  1.94  &  2.45
  \\
J2048  $  -  $  1616  &  3.5  &  61  &   $   $   & $  <26.10^{  }_{  }$   & $  <2.70^{  }_{ }$   & $  <-10.66^{ }_{ }$   & $   <-2.45^{ }_{ }$   &  2.30  &  2.62
  \\ 
J2330  $  -  $  2005  &  4.1  &  76  &   $   $   & $  <-26.56^{  }_{ }$   & $  <2.75^{  }_{ }$   & $  <-11.00^{ }_{ }$   & $   <-2.81^{ }_{ }$   &  <2.18  &  2.45
  \\ 
J2346  $  -  $  0609  &  4.1  &  135  &   $  +  $   & $  -25.13^{ + 0.11}_{ - 0.14}$   & $  4.79^{ + 0.11}_{- 0.14}$   & $  -9.57^{+ 0.11}_{- 0.14}$   & $   -1.52^{+ 0.11}_{- 0.14}$   &  2.99  &  2.09  \\ 

    \end{tabular}}
\end{table*}

\begin{table*}
  \centering
  \caption{Results over a 4~yr timespan for the 16 pulsars with $T>10$~yr from Table~2.}
  \label{results3}
  \renewcommand{\arraystretch}{1.42}
\resizebox{\textwidth}{!}{
    \begin{tabular}{lcccccccccc}
	Name & $T$ & $\overline{\sigma_{\rm W}}$ & log($\overline{{\sigma_{\rm TN}}}$) & log$\overline{|\ddot{\nu}|}$ & log$|n|$ & log($\sigma_z$) & log($\Delta_8$) & log($\sigma_{\rm M}$)  & $\alpha_{\rm TN}$ & $\alpha_z$ \\ 
	& (yr) & ($\mu$s) & (/$\mu$s) & (/s$^{-3}$) &&&& (/$\mu$s) \\ \hline
	J0034  $  -  $  0721  &  4  &  206  &  3.25  & $  <-24.14^{  }_{  }$   & $  <6.55^{  }_{  }$   & $  <-8.58^{  }_{  }$   & $   <-0.64^{  }_{  }$   & $   1.66^{}_{}$   & $  0.14 $  & $  
  $ \\ 
J0108  $  -  $  1431  &  4  &  473  &  <2.98  & $  <-24.80^{  }_{  }$   & $  <7.15^{  }_{  }$   & $  <-9.24^{  }_{  }$   & $   <-1.36^{  }_{  }$   & $   1.00^{}_{}$   & $  $  & $  
  $ \\ 
J0134  $  -  $  2937  &  4  &  27  &  <1.73  & $  <-25.79^{  }_{  }$   & $  <3.83^{  }_{  }$   & $  <-10.24^{  }_{  }$   & $   <-3.13^{  }_{  }$   & $   1.86^{}_{}$   & $  $  & $  
  $ \\ 
J0152  $  -  $  1637  &  4  &  60  &  <2.08  & $  -25.93^{ + 0.18}_{ - 0.29}$   & $  3.61^{ + 0.18}_{ - 0.29}$   & $  -10.37^{ + 0.18}_{ - 0.29}$   & $   -2.48^{ + 0.18}_{ - 0.29}$   & $   2.21^{}_{}$   & $  $  & $  -0.83
  $ \\ 
J0206  $  -  $  4028  &  4  &  201  &  <2.60  & $  <-25.41^{  }_{  }$   & $  <3.83^{  }_{  }$   & $  <-9.86^{  }_{  }$   & $   <-2.09^{  }_{  }$   & $   2.31^{}_{}$   & $  $  & $  
  $ \\ 
J0401  $  -  $  7608  &  4  &  167  &  <2.52  & $  -25.00^{ + 0.05}_{ - 0.05}$   & $  3.84^{ + 0.05}_{ - 0.05}$   & $  -9.44^{ + 0.05}_{ - 0.05}$   & $   -1.73^{ + 0.05}_{ - 0.05}$   & $   2.49^{}_{}$   & $  $  & $  1.37
  $ \\ 
J0452  $  -  $  1759  &  4  &  39  &  2.36  & $  -24.98^{ + 0.10}_{ - 0.14}$   & $  2.71^{ + 0.10}_{ - 0.14}$   & $  -9.43^{ + 0.10}_{ - 0.14}$   & $   -1.72^{ + 0.10}_{ - 0.14}$   & $   3.06^{}_{}$   & $  2.57 $  & $  0.78
  $ \\ 
J0536  $  -  $  7543  &  4  &  144  &  2.47  & $  -25.60^{ + 0.13}_{ - 0.19}$   & $  5.16^{ + 0.13}_{ - 0.19}$   & $  -10.05^{ + 0.13}_{ - 0.19}$   & $   -1.98^{ + 0.13}_{ - 0.19}$   & $   1.67^{}_{}$   & $  0.41 $  & $  -1.07
  $ \\ 
J0630  $  -  $  2834  &  4  &  158  &  3.04  & $  -25.13^{ + 0.16}_{ - 0.23}$   & $  3.45^{ + 0.16}_{ - 0.23}$   & $  -9.57^{ + 0.16}_{ - 0.23}$   & $   -1.51^{ + 0.16}_{ - 0.23}$   & $   2.76^{}_{}$   & $  1.71 $  & $  1.03
  $ \\ 
J0729  $  -  $  1836  &  4  &  92  &  3.66  & $  -23.66^{ + 0.05}_{ - 0.06}$   & $  2.91^{ + 0.05}_{ - 0.06}$   & $  -8.11^{ + 0.05}_{ - 0.06}$   & $   -0.43^{ + 0.05}_{ - 0.06}$   & $   3.61^{}_{}$   & $  3.11 $  & $  1.24
  $ \\ 
J0738  $  -  $  4042  &  4  &  21  &  3.29  & $  -23.97^{ + 0.11}_{ - 0.14}$   & $  4.48^{ + 0.11}_{ - 0.14}$   & $  -8.41^{ + 0.11}_{ - 0.14}$   & $   -0.87^{ + 0.11}_{ - 0.14}$   & $   2.62^{}_{}$   & $  2.19 $  & $  0.88
  $ \\ 
J1456  $  -  $  6843  &  4  &  32  &  2.22  & $  <-24.03^{  }_{  }$   & $  <6.24^{  }_{  }$   & $  <-8.47^{  }_{  }$   & $   <-1.08^{  }_{  }$   & $   1.65^{}_{}$   & $  -0.13 $  & $  
  $ \\ 
J1705  $  -  $  1906  &  4  &  25  &  2.53  & $  -25.11^{ + 0.14}_{ - 0.22}$   & $  2.08^{ + 0.14}_{ - 0.22}$   & $  -9.56^{ + 0.14}_{ - 0.22}$   & $   -2.11^{ + 0.14}_{ - 0.22}$   & $   3.21^{}_{}$   & $  1.15 $  & $  
  $ \\ 
J1717  $  -  $  4054  &  4  &  100  &  <3.51  & $  -24.36^{ + 0.09}_{ - 0.12}$   & $  4.36^{ + 0.09}_{ - 0.12}$   & $  -8.81^{ + 0.09}_{ - 0.12}$   & $   -0.89^{ + 0.09}_{ - 0.12}$   & $   2.63^{}_{}$   & $  1.90 $  & $  0.71
  $ \\ 
J1845  $  -  $  0434  &  4  &  117  &  3.68  & $  -23.60^{ + 0.04}_{ - 0.05}$   & $  3.49^{ + 0.04}_{ - 0.05}$   & $  -8.05^{ + 0.04}_{ - 0.05}$   & $   -0.39^{ + 0.04}_{ - 0.05}$   & $   3.34^{}_{}$   & $  1.70 $  & $  -0.29
  $ \\ 
J1900  $  -  $  2600  &  4  &  54  &  2.36  & $  <-25.38^{  }_{  }$   & $  <5.38^{  }_{  }$   & $  <-9.83^{  }_{  }$   & $   <-2.07^{  }_{  }$   & $   1.55^{}_{}$   & $  1.23 $  & $   $ \\ 
\\[+1em]
	\renewcommand{\tabcolsep}{40mm}
    \end{tabular}}
\end{table*}

\section{Discussion}
There are a number of pulsars with high values of $\sigma_{\rm TN}$ yet with only an upper limit on $\ddot{\nu}$. These pulsars therefore have residuals which are not dominated by a cubic, and this is particularly noticeable in the pulsars with long time-spans. Examples are PSRs~J0034$-$0721, J0536$-$7543, J1456$-$6843 and  J1705$-$1906. The residuals in the first 3 pulsars are dominated by high-frequency noise, with ToAs close together in time showing large offsets compared to the error bars. This is most likely due to profile variability (two of these pulsars are known to be mode-changing). The post-fit residuals for PSR~J1705$-$1906 are shown in Figure~\ref{fig:1705}. The absence of a significant cubic term perhaps indicates that the behaviour of the timing noise is not power-law like in this pulsar (see also \citealt{hlk10}).

For the 88 pulsars for which we measured a significant value of $\ddot{\nu}$, 51 have positive values and 37 negative values. If $\ddot{\nu}$ were simply a measure of the cubic term in the residuals, one would expect equal numbers of positive and negative values. As \citet{jg99} pointed out, if glitch recovery from an unseen glitch in the past dominates the residuals then positive values of $\ddot{\nu}$ are expected. The slight bias towards positive values in our sample, with less than a 20\% probability of obtaining 51 positive values by chance, likely implies that at least some of the pulsars have had glitches in the not too distant past.

\begin{figure}
    \centering
    \includegraphics[width=0.45\textwidth]{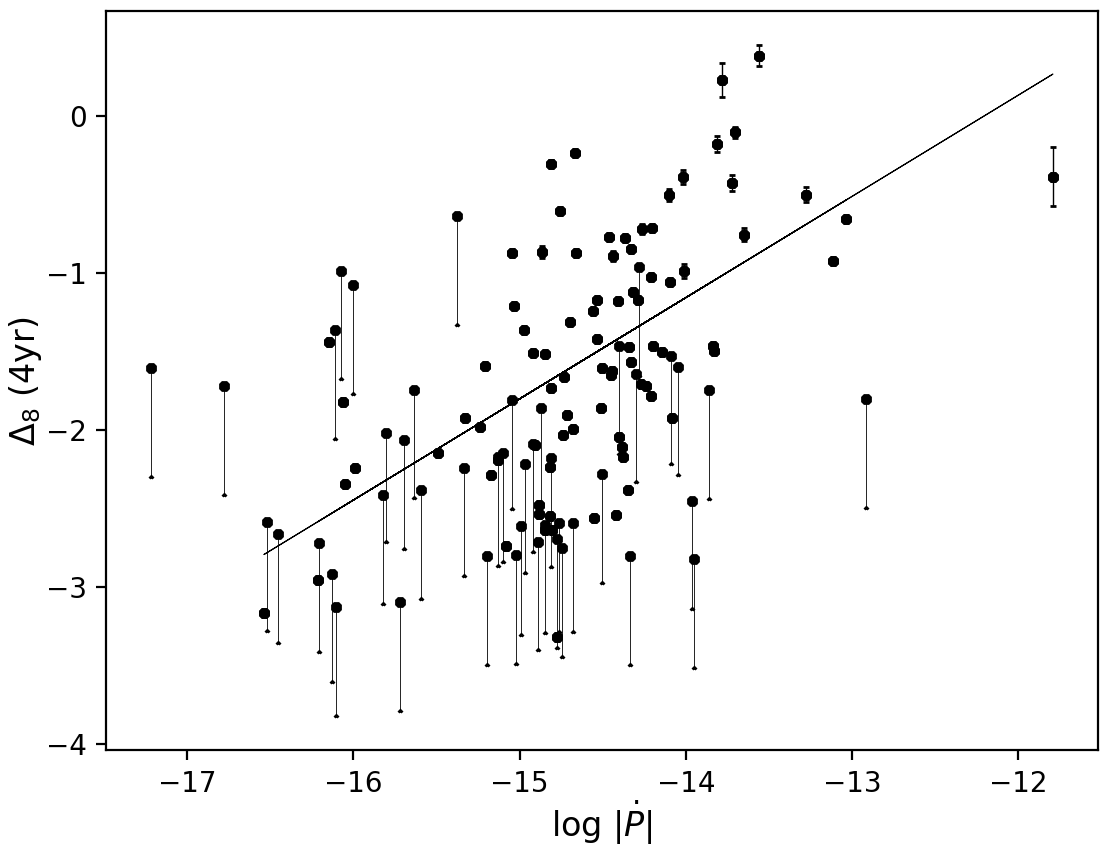}
    \caption{$\Delta_8$ versus $\dot{P}$ for our sample of 129 pulsars. Downward lines with arrows denote objects with upper limits only.}
    \label{fig:del8}
\end{figure}
Figure~\ref{fig:del8} shows $\Delta_8$ versus the pulsar period derivative, $\dot{P}$, for our sample of pulsars. Note that we use values of $\Delta_8$ for $T=4$~yr in order to provide a uniform sample. A straight line fit to the data (without the upper limits included) yields a slope of 0.64 and a correlation coefficient ($cc$) of 0.59.  \citet{antt94} and \citet{hlk10} also found a strong correlation between $\Delta_8$ and $\dot{P}$, with similar values of $cc$ and the slope. As was noted in the introduction, $\Delta_8$ cannot easily be extended for use in pulsars with different values of $T$. This implies that for a homogeneous data set such as we have here, $\Delta_8$ provides a good measure of timing noise but it is not trivial to compare between data sets or to compare pulsars with different observational time-spans.

\begin{figure}
    \centering
    \includegraphics[width=0.45\textwidth]{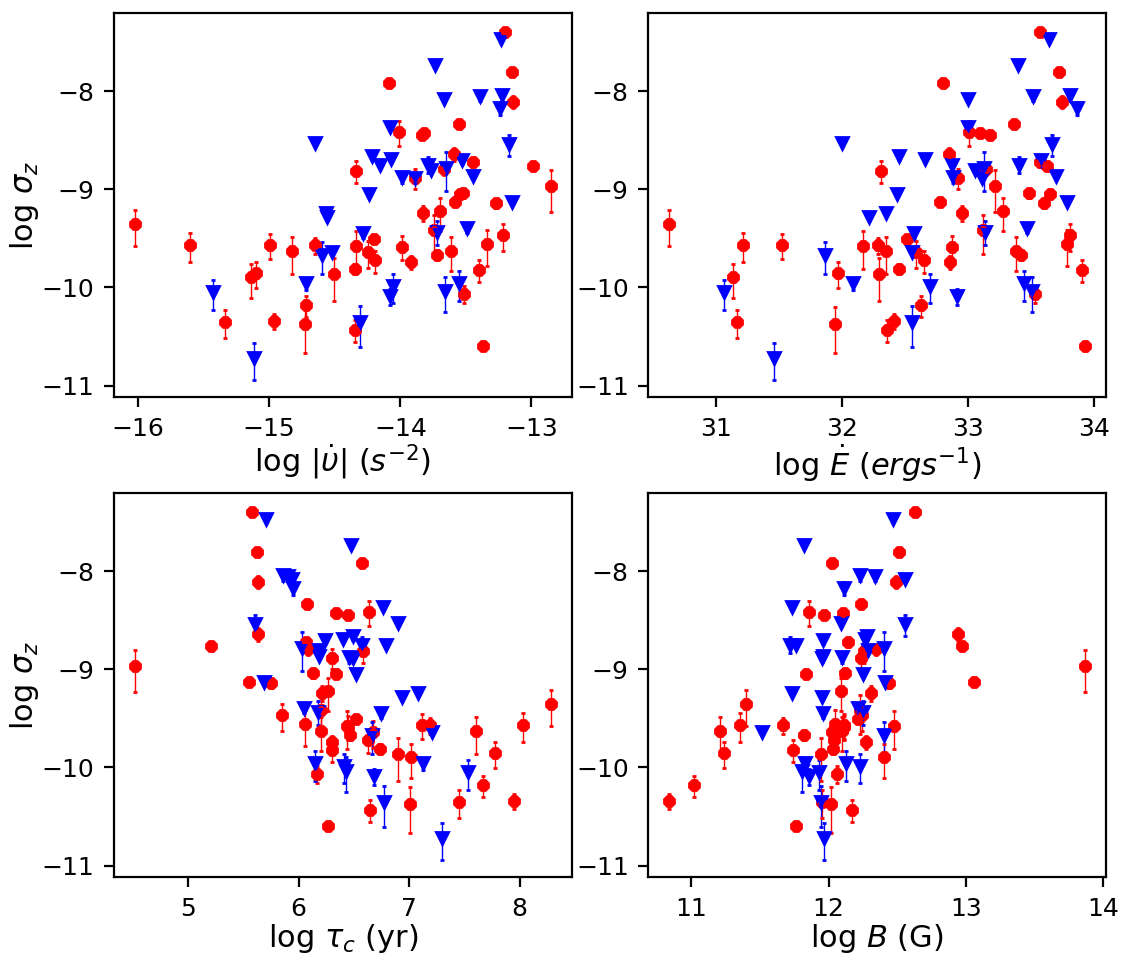}
    \caption{$\sigma_z$ versus $\dot{\nu}$ (top left), $\dot{E}$ (top right), $\tau_c$ (bottom left) and $B$ (bottom right) with a linear fit to the data shown. Pulsars with positive values of $\ddot{\nu}$ are shown as blue triangles, those with negative values as red circles.}
    \label{fig:sigz}
\end{figure}

\begin{figure}
    \centering
    \includegraphics[width=0.45\textwidth]{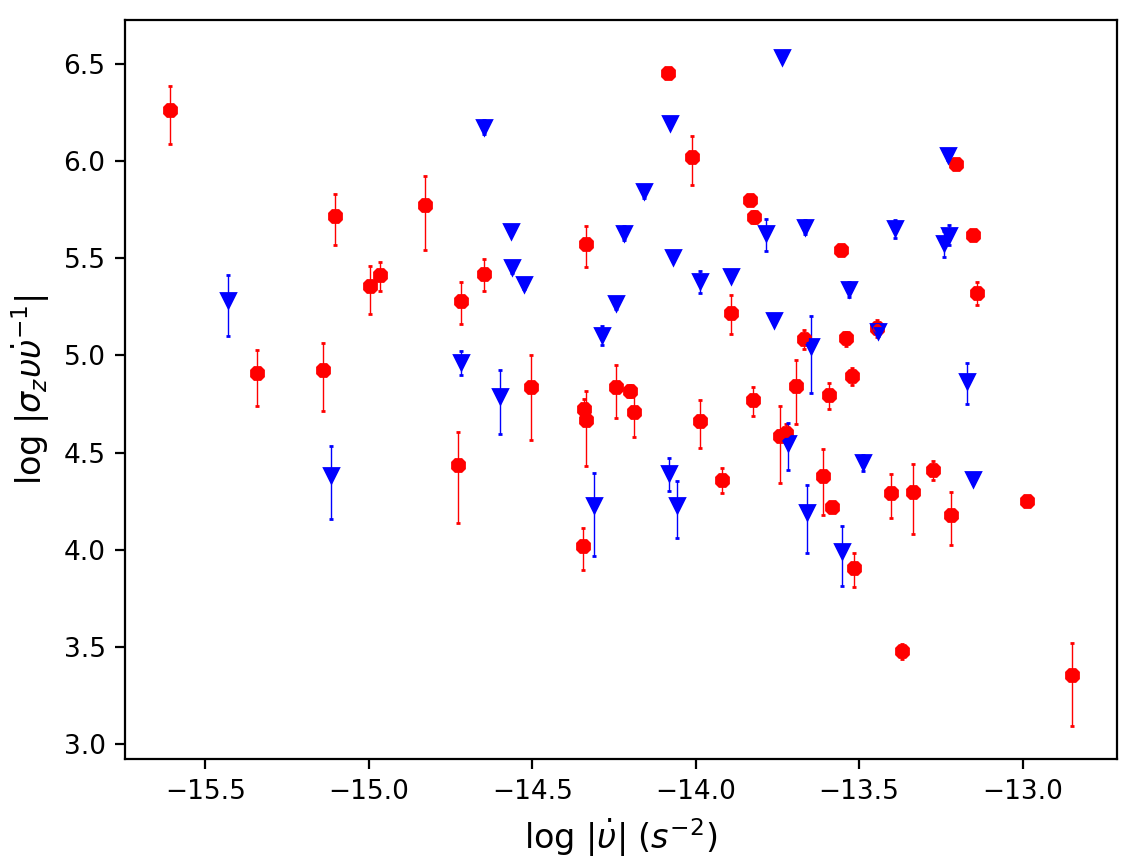}
    \caption{The parameter $\sigma_z\nu\dot{\nu}^{-1}$ versus $\dot{\nu}$. The correlation is much less marked than the equivalent (top left) plot in Figure~\ref{fig:sigz}.}
    \label{fig:mel}
\end{figure}

Figure~\ref{fig:sigz} shows $\sigma_z$ versus various pulsar parameters, again using the $T=4$~yr sample. In this figure we only include those pulsars for which the significance of $\ddot{\nu}$ is more than 2-$\sigma$. As expected, there is a correlation between $\sigma_z$ and $\dot{\nu}$ ($cc=0.51$) though this is weaker than that found by \citet{hlk10} as is the $cc=-0.52$ determined between $\sigma_z$ and $\tau_c$. According to \citet{ml14}, the quantity $\sigma_z \nu \dot{\nu}^{-1}$ should be independent of $\dot{\nu}$. This relationship is shown in Figure~\ref{fig:mel}; the correlation coefficient is $-0.31$ and there remains a slight tendency for pulsars with lower $\dot{\nu}$ to have higher $\sigma_z \nu \dot{\nu}^{-1}$. There is a large spread in values for a given $\dot{\nu}$. \citet{ml14} attribute this spread to variations in the degree of turbulence in the interior fluid of the neutron star. We can also investigate the time dependence of $\sigma_z$ by comparing the $T=4$~yr result in Table~3 with the longer time-spans from Table~2. The exponent of the time dependence, $\alpha_z$, is listed in Table~3. Excluding PSR~J0536$-$7543, the mean value of $\alpha_z$ is 1.0, which from equation~\ref{zpt} implies $\alpha_{\rm R} = -5$ for the power spectral density of the residuals. This value of $\alpha_{\rm R}$ is consistent with the results of \citet{psj+19} who found a mean value for $\alpha_{\rm R}$ of $-5.2$. For three of the brightest pulsars in our sample we can further investigate $\sigma_z$ as a function of $T$ by subdividing the data into 1~yr segments (due to our monthly sampling it is not possible to get shorter segment lengths). This confirms the shallow value of $\alpha_z$ for PSR~J1845$-$0434, the moderate value for PSR~J0738$-$4042 and the large $\alpha_z$ for PSR~J0729$-$1836, again in line with the spread of values seen by \citet{psj+19}.

\begin{figure}
    \centering
    \includegraphics[width=0.45\textwidth]{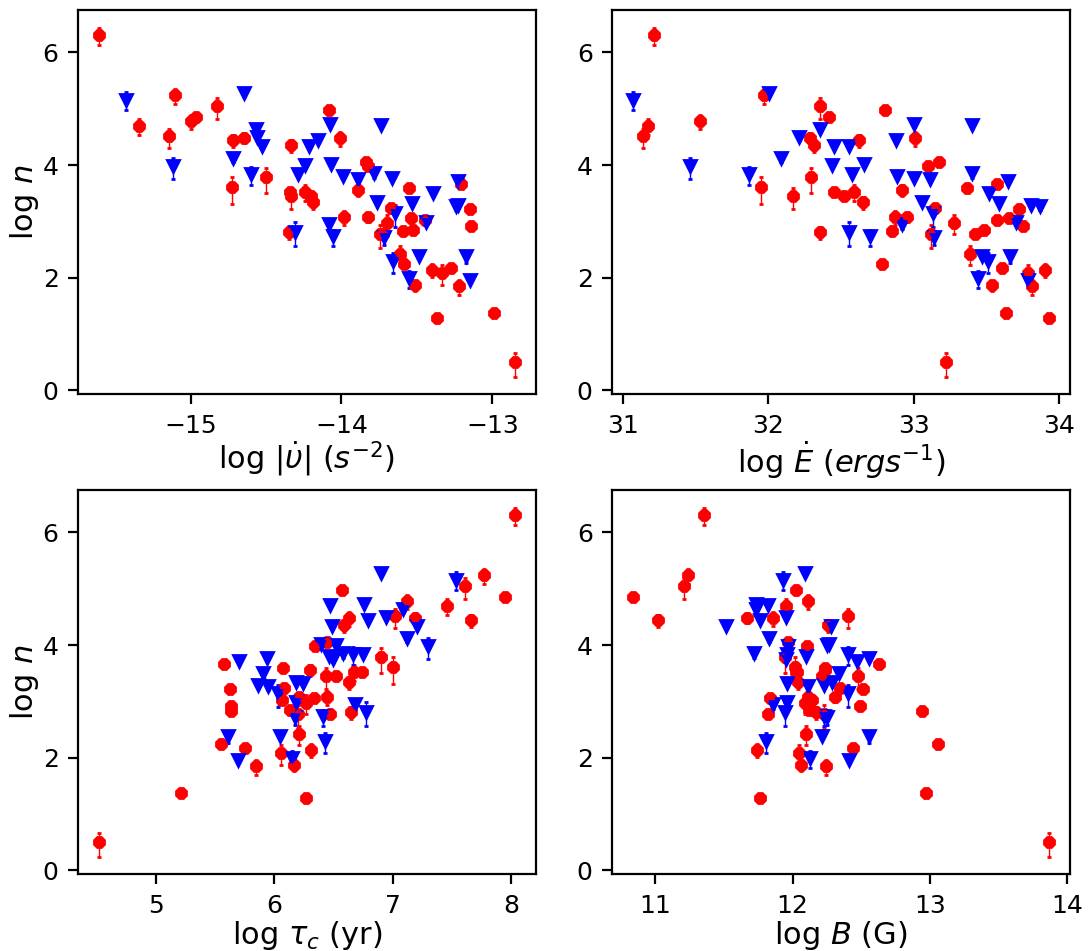}
    \caption{Braking index, $n$, versus $\dot{\nu}$ (top left), $\dot{E}$ (top right), $\tau_c$ (bottom left) and $B$ (bottom right) with a linear fit to the data shown. Pulsars with positive values of $\ddot{\nu}$ are shown as blue triangles, those with negative values as red circles.}
    \label{fig:n}
\end{figure}

Figure~\ref{fig:n} shows $n$ versus the pulsar parameters. Generally there is a very strong correlation between $n$ and the pulsar parameters with $cc=-0.80$ with $\dot{\nu}$, $cc=0.80$ with $\tau_c$ and $cc=-0.71$ with $\dot{E}$. $n$ can therefore be used as a metric for timing noise, but as $n$ has no obvious scaling as a function of $T$, it cannot be compared easily across data sets. It can be seen that the values for $n$ in Table~3 are (in all cases) greater than the values of $n$ in the longer data spans from Table~2. It should also be noted that ranking pulsars via $n$ gives a very different set of objects to ranking on $\sigma_z$ or $\sigma_{TN}$.

\begin{figure}
    \centering
    \includegraphics[width=0.45\textwidth]{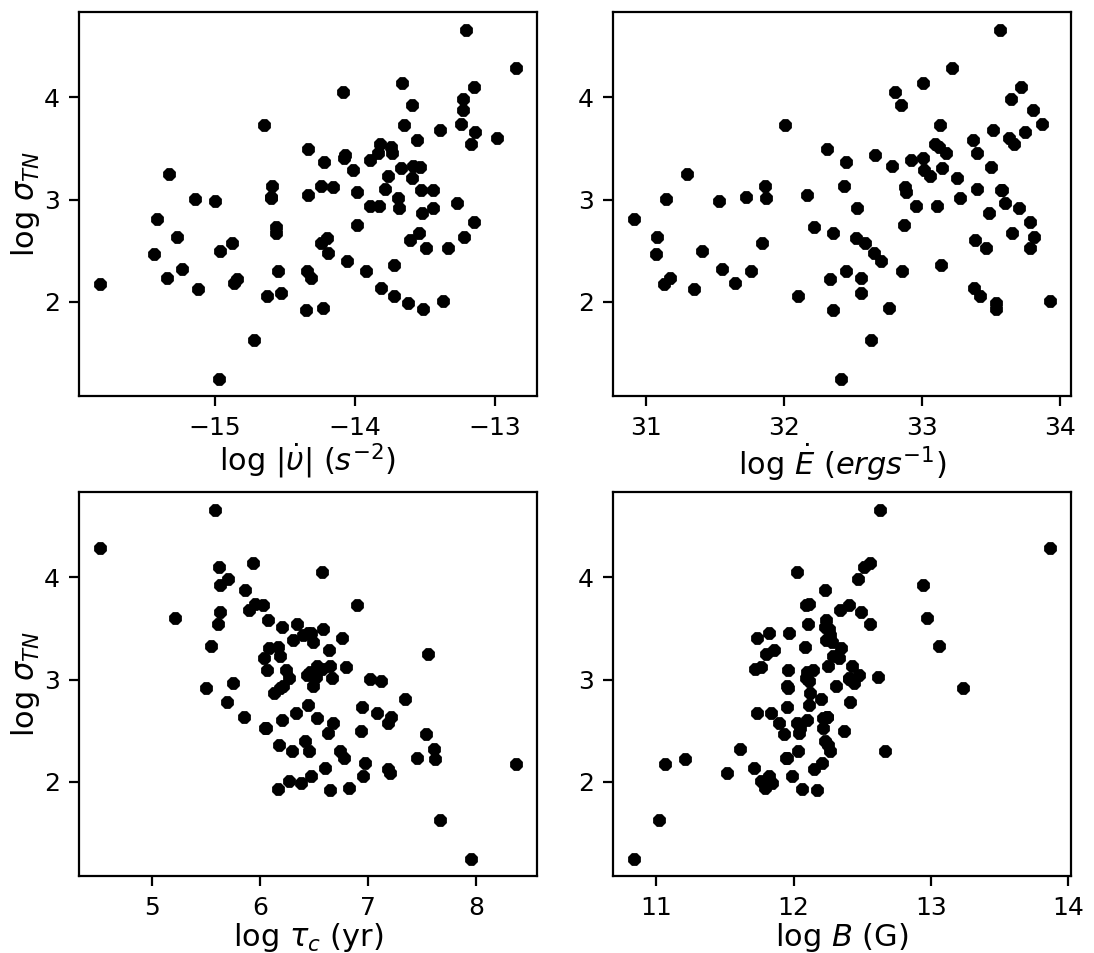}
    \caption{$\sigma_{\rm TN}$ versus $\dot{\nu}$ (top left), $\dot{E}$ (top right), $\tau_c$ (bottom left) and $B$ (bottom right) with a linear fit to the data shown.}
    \label{fig:sigTN}
\end{figure}

Figure~\ref{fig:sigTN} shows $\sigma_{\rm TN}$ versus the pulsar parameters for 91 pulsars with significant values. Again the correlation coefficients between $\sigma_{\rm TN}$ and the pulsar parameters are weak. We can investigate the time dependence of $\sigma_{\rm TN}$ by comparing the $T=4$~yr result with the longer time-span data. The exponent of the time dependence, $\alpha_{\rm TN}$ is given in Table~3. The expectation from \citet{sc10} is that the value of $\alpha_{\rm TN}$ should be 1.9. Although there is a large spread in $\alpha_{\rm TN}$, the mean of the values is indeed 1.9 after removing the three pulsars with non-cubic residuals discussed earlier. 

\begin{figure}
    \centering
    \includegraphics[width=0.45\textwidth]{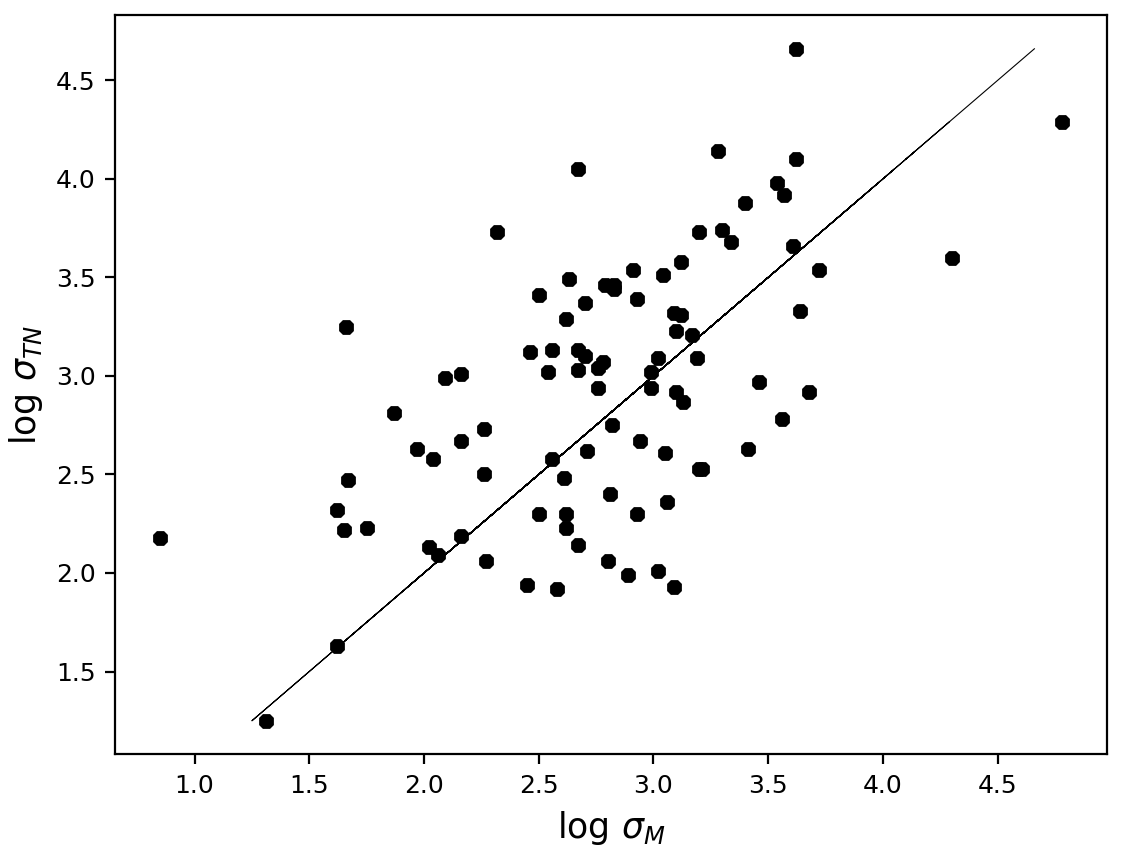}
    \caption{$\sigma_{\rm TN}$ versus $\sigma_{\rm M}$. The straight line shows the 1:1 correspondence.}
    \label{fig:model}
\end{figure}

Figure~\ref{fig:model} shows $\sigma_{\rm TN}$ versus the best model fit from \citet{sc10} given by $\sigma_{\rm M}$ in Tables~2 and 3. Although the correspondence between the model and the data is good, there is a large spread of observed values around a given $\sigma_{\rm M}$, typically of the order of 1.5 dex. This was also recognised by \citet{sc10} and incorporated into their parameter $\delta$. We attempted our own fit to equation~\ref{sigmam}, in a similar way to that performed by \citet{psj+19}. We find a maximum correlation coefficient of 0.62 for $a=-1.7$, $b=1.0$. This is a much steeper dependence on $\nu$ than found by either \citet{hlk10} or \citet{sc10}, and, given that $\dot{P} = \nu^{-2} \dot{\nu}$, shows that $\sigma_{\rm TN}$ also depends much more strongly on $\dot{P}$ than any other pulsar parameter.

This has implications for the potential detectability of gravitational waves via the timing of millisecond pulsars. Let us take a typical old pulsar from our sample, PSR~J0206$-$4028 which has $\nu=1.59$~Hz and $\dot{\nu}=-3.0\times10^{-15}$~s$^{-2}$. After 10~yr of timing, $\sigma_{\rm TN}$ is 1500~$\mu$s. We can scale $\sigma_{\rm TN}$ to a typical millisecond pulsar with $\nu=200$~Hz and $\dot{\nu}=-1.0\times10^{-15}$~s$^{-2}$, via equation~\ref{sigmam}. Using the scaling relation of \citet{sc10}, we predict $\sigma_{\rm M}$ of 6~$\mu$s, whereas using our scaling relationship which has a much steeper dependence on $\nu$ we obtain 0.14~$\mu$s. This is much closer to the observed values, which are typically below 1~$\mu$s for the majority of the millisecond pulsar population and below 0.2~$\mu$s for about 50\% of the sample \citep{srl+15,lcc+17}. Indeed, \citet{lcc+17} also conclude that the dependence on $\nu$ is much steeper than suggested by \citet{sc10}. We therefore conclude that timing noise remains a factor in millisecond pulsar timing, but less so than once feared. In addition, the large spread in timing noise for given pulsar parameters should mean that there exists some millisecond pulsars with extremely low levels of timing noise.


\section{Summary}
We have used regular observations of a group of 129 mainly middle-aged pulsars with the Parkes radio telescopes to measure the timing noise properties of this class of pulsar. We have examined four different metrics for measuring timing noise. Three of these, $\Delta_8$, $\sigma_z$ and $n$ rely on the measurement of $\ddot{\nu}$ using the timing package {\sc tempo2}. The fourth, $\sigma_{\rm TN}$, merely measures the rms residuals after fitting only for $\nu$ and $\dot{\nu}$. All the metrics depend on the time-span of the observations in a way that is difficult to quantify and which is likely to be different for different pulsars. We have shown that $n$ gives the tightest correlation with other pulsar parameters, particularly $\dot{\nu}$. The parameterisation, $\sigma_{\rm M}$ given in \citet{sc10} gives a reasonable fit to the data although there is a significant spread in $\sigma_{\rm TN}$ for any given $\sigma_{\rm M}$. This can be attributed to different nuclear conditions within the different stars as proposed by \citet{ml14} and possibly for evidence for superfluid turbulence in the stellar interior. Finally, in spite of the simplicity in measurement, all of these metrics appear flawed in relying directly on fluctuations in the ToA residuals. A more robust approach would be to model the timing noise as e.g. a red noise process and directly fit for the model along with all the other timing parameters. A full-blown Bayesian method, which incorporates such an approach has been used by \citet{lsc+16} for millisecond pulsars and by \citet{psj+19} for young pulsars. Although computationally expensive, it appears to provide the best mechanism for quantifying timing noise directly.

\section*{Acknowledgements}
The Parkes radio telescope is part of the Australia Telescope, which is funded by the Commonwealth Government for operation as a National Facility managed by CSIRO. NN and PJ received support from DFAT Grant AAC068. SJ thanks NARIT for their hospitality.




\bibliographystyle{mnras}
\bibliography{ben} 


\bsp	
\label{lastpage}
\end{document}